\documentclass[12pt]{article}
\usepackage{geometry} 
\usepackage{graphicx}
\usepackage[parfill]{parskip}
\usepackage{amssymb}
\usepackage{epstopdf}

\title{New expressions for  gravitational scattering amplitudes}
\author{Andrew Hodges\thanks{ andrew.hodges@wadh.ox.ac.uk, http://www.twistordiagrams.org.uk. The author had the honour of presenting this material on 22 July 2011 to a conference at Oxford University celebrating the imminent eightieth birthday of Sir Roger Penrose, FRS, OM.}\\{\footnotesize  {\it Wadham College, University of Oxford, Oxford OX1 3PN, U.K. }}}
\date{25 August 2011}
\begin{document}
\maketitle
\begin{abstract}
New methods are introduced for the description and evaluation of tree-level gravitational scattering amplitudes. An N=7 super-symmetric recursion, free from spurious double poles, gives a more efficient method for evaluating MHV amplitudes. The recursion is naturally associated with twistor geometry, and thereby gives a new interpretation for the amplitudes. The recursion leads to new expressions for the MHV  amplitudes for six
 and seven gravitons, simplifying their symmetry properties, and suggesting further generalization. The N=7 recursion is valid for all tree amplitudes, and we illustrate it with a simplified expression for the six-graviton NMHV amplitude. Further new structure emerges when MHV amplitudes are expressed in terms of momentum twistors.

\end{abstract}

\section{Introduction}

Despite many recent advances, it remains difficult to evaluate and express even the simplest gravitational amplitudes in a manner which reflects their symmetries. Even for MHV amplitudes, there is nothing comparable to the simplicity of the Parke-Taylor formula in pure gauge theory. The recent survey of Nguyen, Spradlin, Volovich and Wen (2009) gives a vivid account of the problem, and has helped to motivate the new approaches suggested in this note.  

The discussion that follows is founded on the BCFW recursion procedure (Britto, Cachazo, Feng and Witten, 2005). The first new idea introduced is the employment of an ${\rm N=7}$ super-symmetric formalism in BCFW recursion, rather than the standard ${\rm N=8}$. This greatly improves the effectiveness of the recursion, eradicating spurious double poles. It yields a very simple derivation of the BGK formula (Berends, Giele and Kuijff, 1988)  for all MHV amplitudes, which has been more recently confirmed by Mason and Skinner (2008). We  relate this also to the KLT  formula (Kawai, Lewellen and Tye, 1986), as applied to the special MHV case. But the N=7 recursion also leads to a quite new form for the 6- and 7-graviton MHV amplitudes, with greater symmetry, and it suggests that these expressions will generalise to a new normal form for all $n$. A simpler structure emerges for general tree-level amplitudes. We illustrate this by deriving an expression for the 6-field NMHV amplitude.\footnote{Version 1 of this paper, published on 10 August 2011, lacked a demonstration that the recursion was valid beyond MHV amplitudes. This Version 2 includes a new section 13 which repairs this gap, thanks to a contribution from H. Elvang.}

In a further development, we write MHV amplitudes in terms of the {\em momentum-twistors} introduced in (Hodges 2009). This is a useful innovation even though there is no apparent r\^{o}le for the dual conformal invariance that momentum twistors were devised to represent. The emergent feature, proved for $n \le 6$, is that a numerator polynomial in the momentum-twistors captures the content of the gravitational amplitude.

This last development is explicitly concerned with giving twistor representation, but in fact the original idea of the N=7 calculus was also inspired by twistor geometry and we shall motivate it by using the twistor-geometric version of BCFW recursion. This reflects an underlying principle that twistor geometry will be useful in eliciting new representations.  In particular, twistor representation of a structure which {\em breaks} conformal symmetry, as gravity does, can make that broken symmetry manifest in a new and useful way.  The essential point is that  {\em linearized} gravity only breaks conformal invariance in a very weak sense, introducing non-singular {\em numerator} factors, while the underlying singularity structure depends only on conformally invariant geometry. We shall exploit the simplicity of these numerators. Historically,  Roger Penrose initiated the twistor description of gravity over 40 years ago, and the gravitational scattering diagrams in (Penrose and McCallum 1972), since neglected, embody the essential concepts used in this analysis.

This approach is thus different from the currently prevailing line of attack on gravitational amplitudes, which is based on extending the insights from string theory which led to the KLT relations. A recent paper (Carrasco and Johansson 2011) gives copious references to this line of development, which is of great significance for computing loop integrals in N=8 super-gravity. This approach, as currently formulated, is much less natural in twistor geometry. Indeed there is a certain complementarity here. String-based relations make a complete break from conformally invariant geometry but give rise to powerful relations between amplitudes which are independent of helicity. Twistor-based relations emphasise the connection with conformal geometry, but at the cost of according what may seem an excessively dominant r\^{o}le to helicity conservation and violation. Hopefully, further developments  will before long unify these approaches, perhaps by making use of  insights from Witten's twistor string model.

\section{Three-point amplitudes}

We first review the {\em twistor diagram} formalism as originated by Penrose (1972) and developed by this author for gauge-field scattering. The central point is that the material in (Hodges 2005a) showed how the original BCFW recursion is equivalent to a simple and  natural composition of twistor diagrams. The super-symmetric extension of the diagram formalism (Hodges 2005b) then showed its computational value (Hodges 2006). 

The diagrams are not essential to the new formulas for amplitudes to be arrived at in section 10,  but they assist visualisation of the various symmetries of the theory, and cast new light on the resulting expressions. In particular, they make contact with the significant theme of representing amplitudes as {\em integrals over geometric regions}, so that addition formulas arise from nothing but the addition of regions. This emergent geometric concept has played a useful part in recent advances, beginning with (Hodges 2009). Another reason for interest in the diagram formalism is that it is closely connected the {\em  Grassmannian} description of amplitudes.

Indeed Nima Arkani-Hamed, Freddy Cachazo and their group discovered the twistor Grassmannian structure for the amplitudes  after studying the properties of twistor diagrams (Arkani-Hamed, Cachazo, Cheung and Kaplan 2009a, 2009b). Development of the Grassmannian calculus has since then inspired entirely new insights.  So far it has been defined only where there is a given ring-ordering of the fields, as in the colour-ordered sectors of the (planar part of) gauge theory. But current investigations (Arkani-Hamed 2011) suggest the existence of a more fundamental picture in which different orderings can be combined. One would expect this to encompass the various relations (Kleiss-Kuijff, etc.) between amplitudes associated with different colour-orderings, and perhaps also gravitational amplitudes. Thus formulating gravitational amplitudes as twistor diagrams should  make a useful contribution to such investigations. 
\newpage
The starting-point for BCFW recursion lies in the 3-field amplitudes, and it is the  simple 3-field amplitude for gravity which makes the recursive principle such an attractive proposition. We shall first fill in some material stated but not given a detailed explanation in (Hodges 2005a, 2005b). 

Twistor diagram structure is based on the twistor-integral representation of the momentum delta-function on massless fields, i.e.\ the operation of integrating the product of some set of massless fields over Minkowski space. For {\em four} massless scalar fields, the representation is particularly simple, being conformally invariant, and can be written as the diagram:
\begin{figure}[h] 
   \centering
   \includegraphics[width=124px]{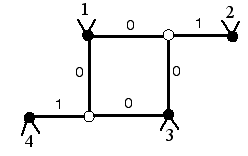} 
  \end{figure}
\vspace{-25mm}
\begin{equation} \label{eq:0000box}\end{equation}
\vspace{10mm}

Here the black and white vertices denote twistor ($Z^{\alpha}$) and dual twistor  ($W_{\alpha}$) variables, all to be integrated out with the natural form. Each thick line represents a pole singularity: the label 0 denotes a simple pole of form $(W.Z)^{-1}$, and the label 1 denotes a double pole of form $(W.Z)^{-2}$. We ignore questions of sign and factors of $2\pi i$ in what follows.\footnote {We are also over-simplifying: for genuine contours to exist for these twistor integrals, the formalism requires a shift from $W.Z=0$ to $W.Z=k$, as described in (Hodges 2005a). But for the developments described in this paper, this refinement can be safely ignored.} The external labels $1, 2, 3, 4$ denote twistor functions corresponding to massless scalar fields, with the two `ears' as a reminder that these functions are actually cohomological. The double poles are examples of twistor transform lines, which do nothing but transform the twistor representative of a field to the dual twistor representative. The single poles can be considered as Cauchy poles which reduce the integral to one where all the integrated-out variables lie on the same line in projective twistor space. The space of such lines is equivalent to the space of points in Minkowski space, and so the complete twistor integral is equivalent to a $\int d^4x$ integration over Minkowski space.

An integral form for {\em three} massless scalar fields is obtained  by taking the fourth scalar field to be the {\em constant} field, i.e.\ $\phi_4(x) \equiv 1$. (A more rigorous treatment would have to consider the exact sense in which this limit is taken, but this is not necessary at this level of formality.) In the twistor picture, the constant field is just the {\em elementary state based at infinity}, with representation $f(Z_4) = \{(Z_4.I_1)(Z_4.I_2)\}^{-1}$, where $I_{1\alpha}$ and $I_{2\beta}$ are two dual twistors whose skew product is the infinity twistor $I_{\alpha \beta}$. 
\newpage
Integrating out $Z_4$ leaves the expression
\begin{figure}[h] 
   \centering
   \includegraphics[width=96px]{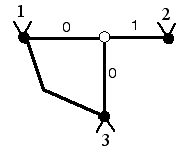} 
  \end{figure}
  \vspace{-25mm}
\begin{equation} \label{eq:phi3}\end{equation}
\vspace{5mm}

as the twistor-diagram version of the scalar $\phi^3$ vertex, i.e.\ the delta-function $1. \delta(\sum_1^3 p_i)$.
Here the crooked leg between 1 and 3 denotes the factor $(I_{\alpha\beta}Z_1^{\alpha}Z_3^{\beta})^{-1}$, where $I_{\alpha\beta}Z_1^{\alpha}Z_3^{\beta}$ is equivalent to the product of spinors conventionally  written $\langle13\rangle$. At first sight this  is an asymmetrical expression, but symmetry is restored by a formal integration by parts. One advantage of the diagram formalism is that identities are readily expressed by operations of  integrating by parts on the integrands, simple because the integrands are so simple.

To translate \begin{equation}\frac{\langle 12\rangle^{3}}{\langle 23\rangle \langle 31 \rangle} \delta(\sum_1^3 p_i)\, ,\end{equation} the standard gauge-theoretic 3-point amplitude, we therefore apply the operators 
\begin{equation}(I_{\alpha\beta}Z_1^{\alpha}Z_2^{\beta})^3 (I_{\alpha\beta}Z_2^{\alpha}Z_3^{\beta})^{-1}(I_{\alpha\beta}Z_3^{\alpha}Z_1^{\beta})^{-1}\label{eq:operators} \end{equation}
 to (\ref{eq:phi3}). Before doing this, we introduce some further notation. Following Penrose (1972, 1975) it is helpful to write the simple pole $(W.Z)^{-1}$ as $[W.Z]_1$, then to define $[W.Z]_n$ in such a way that $\frac{\partial}{\partial Z^{\alpha}} [W.Z]_n = W_{\alpha}[W.Z]_{n+1}$. For positive $n$ this gives a pole of order $n$ together with the factor $(n-1)!$. Thus $[W.Z]_{n+1}$ is the factor which in diagrams is indicated by a  thick line labelled with $n$.

But now we can extend the definition to non-positive integers. For  $n=0$, the formal object $[W.Z]_0$ can be defined as a contour with {\em boundary} on $W.Z=0$. Then $[W.Z]_{-1} $ is defined as the {\em numerator} $(W.Z)$, combined with such a boundary contour, $[W.Z]_{-2} $ as the numerator $\frac{1}{2}(W.Z)^2$ combined with such a boundary, and so on.

In the original Penrose notation the object $[W.Z]_0$ was indicated by a $(-1)$ label, and then later, as in (Hodges 2005a), a {\em wavy} line was used. Typographically this is too challenging;  we are now adopting a single simple {\em thin} straight line to indicate  boundary-defining elements such as $[W.Z]_{0}$, with a {\em double thin} line for $[W.Z]_{-1}$. The reason for this notation is that these boundary-defining elements emerge as the most fundamental objects of the theory, and so should have the simplest possible expression. They express the anti-derivatives involved both in the concept of propagator, and in the concept of gauge-field. 

Now we are ready to apply the operators (\ref{eq:operators}). 
The denominators cancel and the result can be written simply as:
\begin{figure}[h] 
   \centering
   \includegraphics[width=87px]{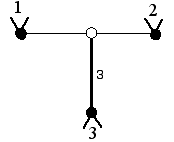} 
  \end{figure}
\vspace{-25mm}
\begin{equation} \end{equation}
\vspace{5mm}

All the results in (Hodges 2005a) came from applying BCFW composition to this primitive object and its dual. 

The super-symmetric generalization in (Hodges 2005b) came from replacing all the lines by N=4 super-boundaries and all the external functions by super-fields.  The 3-field super-amplitude then becomes symmetric in the three external super-fields, as:
\begin{figure}[h] 
   \centering
   \includegraphics[width=85px]{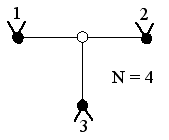} 
  \end{figure}
  \vspace{-25mm}
\begin{equation} \end{equation}
\vspace{5mm}

(The notation in (Hodges 2005b) introduced {\em arcs} instead of straight lines to distinguish the N=4 super-boundaries from the classical boundaries, but now that we need to introduce N=7 and N=8 as well, this distinction will be dropped, and an indication of the super-symmetry given explicitly instead.)

The super-symmetric extension of BCFW is automatic in this formalism, and all amplitudes become characterized as simply integrals of the external fields over regions with boundaries specified by  these elements.

Having rehearsed this derivation of 3-amplitudes in N=4 gauge theory, it is
straightforward to adapt it to find the analogous gravitational 3-amplitude 
\begin{equation}\frac{\langle 12\rangle^{6}}{\langle 23\rangle ^2 \langle 31 \rangle^2}  \delta(\sum_1^3 p_i)\, .\end{equation}
\newpage
We obtain, for linearized gravity without super-symmetry, the 3-vertex:

\begin{figure}[h] 
   \centering
   \includegraphics[width=91px]{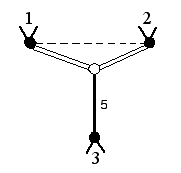} 
  \end{figure}
  \vspace{-25mm}
\begin{equation} \end{equation}
\vspace{5mm}

  where the dashed line indicates a simple {\em numerator} factor $I_{\alpha\beta}Z_1^{\alpha}Z_2^{\beta} =\langle12\rangle$.
It is simple to extend this to super-fields with N=8 super-symmetry, obtaining:

\begin{figure}[h] 
   \centering
   \includegraphics[width=92px]{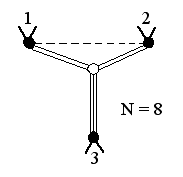} 
  \end{figure}
\vspace{-25mm}
\begin{equation} \end{equation}
\vspace{5mm}

Using integration by parts, this expression is readily seen to be symmetric in $\{123\}$.

At first sight these twistor expressions, which emphasise conformal symmetry,  have lost the simplicity of the statement that `gravity is the square of Yang-Mills'. But they convey the more accurate idea that `gravity times $\phi^3$ is the square of Yang-Mills', which correctly accounts for the {\em square} of the delta-function in the equation. In the diagrams we see the conformal-symmetry-breaking denominator factor in the $\phi^3$ vertex complementing the conformal-symmetry-breaking numerator factor in the gravitational vertex. 
\newpage
\section{The BCFW recursion for N=8 super-gravity}

The use of N=8 super-symmetric BCFW recursion is well established for the calculation of tree amplitudes, as is discussed by  (Drummond, Spradlin,  Volovich and Wen 2009). In particular, it can be employed to find the 4-field (super-)amplitude from the 3-amplitudes.

Pivoting on $(23)$,
we have that $M(1234)= M(13k)\circ M(24k) + M(34k)\circ M(12k)$.  The BCFW composition may be written diagrammatically as:

\begin{figure}[h] 
   \centering
   \includegraphics[width=260px]{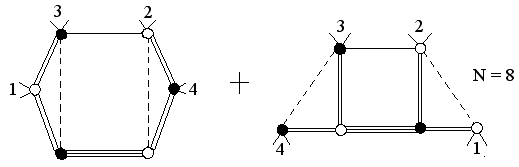} 
  \end{figure}
  \vspace{-25mm}
\begin{equation} \end{equation}
\vspace{5mm}

  The first of these diagrams can be simplified by (i) integration by parts and (ii) eliminating a double twistor transform, giving:
 
 \begin{figure}[h] 
   \centering
   \includegraphics[width=232px]{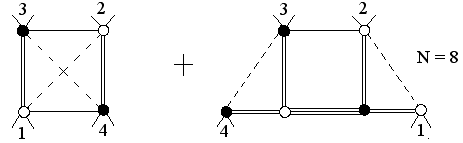} 
  \end{figure}
\vspace{-25mm}
\begin{equation} \label{eq:n8form}\end{equation}
\vspace{5mm}

  This sum is correct for any super-fields, but to check and illustrate the result we shall specialise to the simplest case,
 $M(1^+2^+3^-4^-)$. In this case the super-symmetry is trivially integrated out.  Equivalently, for this particular helicity and pivoting we could have used the  formalism without any super-symmetry. So the diagrams reduce to the N=0 diagrams:
 
 \begin{figure}[h] 
   \centering
   \includegraphics[width=220px]{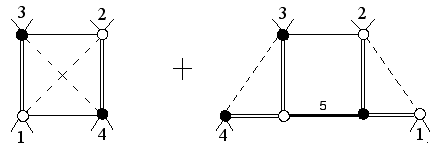} 
  \end{figure}
  \vspace{-25mm}
\begin{equation} \end{equation}
\vspace{5mm}

  where the external functions are now all of classical homogeneity degree $(-6)$.
 
  These diagrams are equivalent to the terms: 
  \begin{equation}\frac{[12]\langle34\rangle^7}{\langle24\rangle\langle13\rangle\langle41\rangle^2\langle23\rangle^2} -
  \frac{[12]\langle34\rangle^6}{\langle12\rangle\langle41\rangle^2\langle23\rangle^2} =\frac{[12]\langle34\rangle^6}{\langle24\rangle\langle13\rangle\langle12\rangle\langle41\rangle\langle23\rangle}\end{equation}
which is the correct amplitude.  

So far the diagrams have added nothing to the results obtained, and indeed this last calculation is completely equivalent to that of Cachazo and  Svr\u{c}ek (2005), with their early use of  BCFW recursion without super-symmetry.

 But some features of the amplitude are well illustrated by the diagrammatic representation. The breaking of conformal invariance by a numerator factor is particularly clear. Momentum conservation is built into the diagram, and identities such as $ [13]\langle24\rangle^{-1} = - [34]\langle12\rangle^{-1}$ amount to simple steps of integrating by parts. The double lines code double poles. But the cancellation of the double poles in the sum remains just as obscure in the diagram formalism as in the standard spinor calculus.

Indeed the diagrams just make more vivid the general inelegance of the N=8 calculus: they are more complex than the result. This drawback becomes more noticeable in calculating more advanced results, where the cancellation of double poles becomes less and less tractable with the growth of algebraic complexity, as is shown explicitly in (Cachazo and  Svr\u{c}ek, 2005). Field theory seems to have made a curiously retrograde step:  it was the triumph of DeWitt (1967) --- in work that established the vital role of the helicity representation, the foundation of all modern developments --- to calculate the 4-graviton amplitude in the form that would now be written as
\begin{equation}\frac{ [12]^4\langle34\rangle^4}{s_{12} s_{13} s_{14}}\, .\label{eq:dewitt} \end{equation}
This formula  brings out the a simplicity and symmetry of the amplitude which is so lacking in this recursion method. What is the origin of the spurious double poles and the asymmetry of the BCFW calculus? 

The answer is that  the momentum conservation condition $s_{12}+s_{13}+s_{14}=0$ has been used to write $(s_{12}s_{13}s_{14})^{-1}$ as $ - s_{14}^{-2}s_{12}^{-1} - s_{14}^{-2}s_{13}^{-1}$. This  splits the amplitude into two terms which each resemble the gauge-theoretic 4-field analogue, with two kinematical denominators rather than three. Such terms have natural representations as box-shaped twistor diagrams, such diagrams being equivalent to the BCFW joining of two 3-vertices. At first sight there might seem to be no other way of  fitting three singularities into a calculus designed to encompass just  two, and so it might appear  that these essentially artificial double poles would be ineradicable within any formalism making use of on-shell recursion methods.

Yet it turns out that these double poles can be eliminated at source by making use of ${\mathrm N=7}$ rather than ${\mathrm N=8}$ supersymmetry. One way of seeing why ${\mathrm N=7}$ is relevant is to observe that although gauge-field amplitudes have genuine fourth-power behaviour, and although gravity is characterized as `the square of Yang-Mills', gravitational amplitudes never actually contain an eighth power. {\em Seventh} powers are the highest that actually occur, and the eighth powers are, like the double poles, spurious.  We shall look at this in more detail shortly, but as a preparatory step look first at a remarkable contour structure that emerges from twistor diagrams, and becomes the key to fitting in the third singularity.

\section{Amplitude identities as contour structures}

The simplest relation between differently colour-ordered amplitudes is that
\begin{equation}A(1234) + A(1243) + A(1423)=0\, ,\end{equation}
which can be read as saying that when a fourth field is added to the ordered three-vertex $A(123)$, the three possible points of insertion yield colour-stripped amplitudes which sum to zero. (Hence, if the gauge field is U(1), the amplitude vanishes, so that this equation can be seen as the simplest case of the `U(1) decoupling identity'.)

If we write this identity in the form of N=4 twistor diagrams, we can draw them in such a way as to emphasise that they give the effect of inserting field 4 in three different ways into the 3-amplitude $A(123)$: 
 \begin{figure}[h] 
   \centering
   \includegraphics[width=309px]{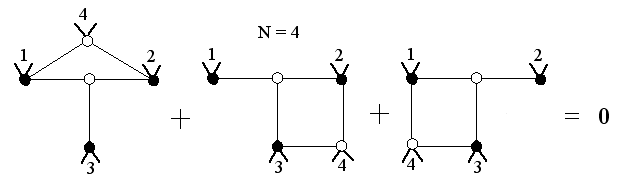} 
  \end{figure}
  \vspace{-25mm}
\begin{equation} \label{eq:3term}\end{equation}
\vspace{5mm}

and this indicates that the zero sum is actually an implication of the more primitive contour identity:

 \begin{figure}[h] 
   \centering
   \includegraphics[width=164px]{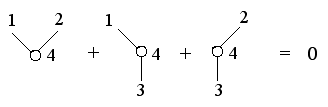} 
  \end{figure}
  \vspace{-25mm}
\begin{equation} \end{equation}
\vspace{5mm}

We shall call this `the triangular contour identity.' This structure has remarkably wide implications. It can be used to express  the U(1) decoupling identity, and more generally the Kleiss-Kuijff relations,  for all MHV amplitudes, these relations also being related to the  validity of  BCFW evaluation with {\em non-adjacent} pivots. But this would go beyond the topic of this note, and here we will confine ourselves to three observations directly relevant to later claims in this paper.

\newpage
(1) The {\em square identity} of N=4 gauge theory follows from the triangular contour identity, taken together with the twistor transform and the dual of the triangular contour identity. To see this, note that the second and third terms in (\ref{eq:3term}) can be seen as the result of inserting the 3 field into the dual 3-amplitude for $\{124\}$. Hence by applying the dual triangular identity, 

 \begin{figure}[h] 
   \centering
   \includegraphics[width=258px]{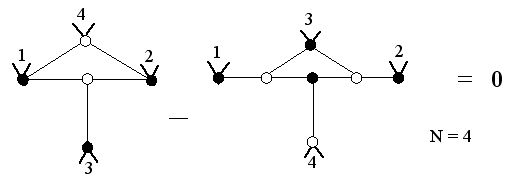} 
  \end{figure}
  \vspace{-25mm}
\begin{equation} \label{eq:squareid}\end{equation}
\vspace{5mm}

which is the square identity, i.e.\ the self-duality of the twistor diagram for the 4-field amplitude.
Hence  all the equivalences of diagrams for N=4 gauge theory, obtained by using the square identity, can be regarded as implications of the triangular contour identity.
 
(2) Now consider the MHV amplitude for {\em five} gauge fields. We may note that the symmetrized sum $A(12345)+A(13245)$ is actually symmetric in $\{235\}$ and in $\{14\}$. That this is so follows straightforwardly by  the calculation
\begin{equation}\frac{1}{\langle12\rangle\langle23\rangle\langle34\rangle\langle45\rangle\langle51\rangle}+\frac{1}{\langle13\rangle\langle32\rangle\langle24\rangle\langle45\rangle\langle51\rangle}= \frac{\langle14\rangle}{\langle12\rangle\langle13\rangle\langle24\rangle\langle34\rangle\langle45\rangle\langle51\rangle}\, .\end{equation}

On the other hand, we can see this symmetry directly and graphically by using diagrams. Attach the 5 field consistently between the 4 and the 1 field in the dual of the 3-term relation (\ref{eq:3term}), using twistor transforms as necessary. We obtain

\begin{figure}[h] 
   \centering
   \includegraphics[width=307px]{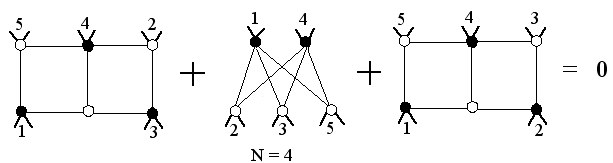} 
  \end{figure}
    \vspace{-25mm}
\begin{equation} \label{eq:5sym}\end{equation}
\vspace{5mm}

The first term is $A(13245)$ and the third term is $A(12345)$. The second term manifestly has the symmetry stated.
We shall find a gravitational analogue of this symmetry in section 8 below.
\newpage
(3) Again starting with the three term identity (\ref{eq:3term}), operate with:
$$(I^{\alpha\beta}W_{4\alpha}\frac{\partial}{\partial Z_3^{\beta}})^{-1} (W_4.Z_2)I^{\gamma\delta}\frac{\partial}{\partial Z_2^{\gamma}}\frac{\partial}{\partial Z_3^{\delta}}\, .$$

The first term vanishes (using the fact that $(W_4.Z_2)[W_4.Z_2]_{1} = 0$), and we obtain a  two-term identity:

\begin{figure}[h] 
   \centering
   \includegraphics[width=275px]{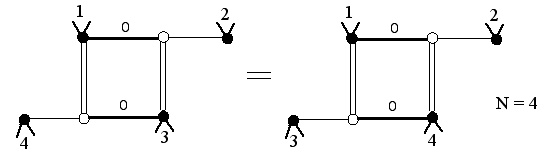} 
  \end{figure}
    \vspace{-25mm}
\begin{equation} \label{eq:twoterm} \end{equation}
\vspace{5mm}

We will call this the `square symmetry identity'. Clearly, this is equivalent to knowing that the scalar box diagram (\ref{eq:0000box}) is completely symmetric in all four fields, so it does not tell us anything new; but it is remarkable that this symmetry can be seen as a consequence of the more fundamental triangular contour identity.
 
 These three examples should suffice to show the importance of this additive contour structure. It is worth noting that such additions of contours naturally take us beyond the scope of the twistor diagrams representing the colour-stripped amplitudes in N=4 gauge theory. Such diagrams behave like topological disks, with the colour-trace running round the exterior as in a ribbon diagram, this property being conserved by the BCFW joining rule. The more general diagrams introduced here are not necessarily planar, as for instance the third term in (\ref{eq:5sym}). This is of course appropriate as we are intending to use this more general type of diagram for gravitational amplitudes, which have no natural planarity, and correspond to closed rather than open strings.

As the triangular contour identity is a purely topological idea, it is not confined to N=4 gauge theory, and this is what we shall next exploit.

\newpage
\section{The 4-field amplitude as one twistor diagram}

We now consider a different helicity choice, namely  $M(1^-2^+3^+4^-)$, for evaluation by means of the general N=8 formula (\ref{eq:n8form}). Using twistor transforms and integration by parts, we write the diagrams in such a way that the external fields for these helicities will all have classical homogeneity $(+2)$, thus:

 \begin{figure}[h] 
   \centering
   \includegraphics[width=239px]{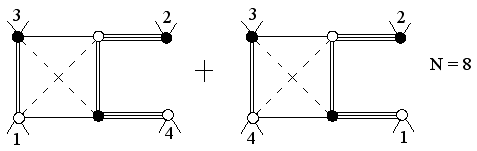} 
  \end{figure}
  \vspace{-25mm}
\begin{equation} \end{equation}
\vspace{5mm}

If we wished we could perform the eight-fold fermionic integration and leave the sum of two N=0 diagrams:

 \begin{figure}[h] 
   \centering
   \includegraphics[width=211px]{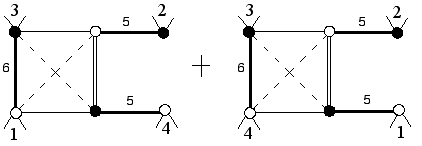} 
  \end{figure}
  \vspace{-25mm}
\begin{equation} \end{equation}
\vspace{5mm}

These then give the correct amplitude as the sum:
\begin{equation}\frac{[14]\langle14\rangle^6}{\langle12\rangle\langle34\rangle\langle23\rangle^3} -
  \frac{[14]\langle14\rangle^6}{\langle24\rangle\langle13\rangle\langle23\rangle^3}=\frac{[14]\langle14\rangle^7}{\langle12\rangle\langle34\rangle\langle13\rangle\langle24\rangle\langle23\rangle^2}\end{equation}

But instead of expanding the fermionic parts  completely, we may
integrate out only the {\em eighth} of  the eight super-components throughout. With this helicity choice there are no super-components on the external fields, and the effect is to leave an integral with {\em seven} super-components still to be integrated:

 \begin{figure}[h] 
   \centering
   \includegraphics[width=240px]{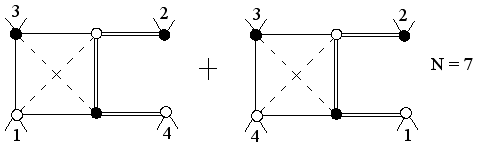} 
  \end{figure}
    \vspace{-25mm}
\begin{equation} \end{equation}
\vspace{5mm}

  in which the crucial change is that the lines from 3 to 1, 3 to 4, have become simple super-boundary lines. This means that they can be combined through the triangular contour identity, into a single diagram:
  \newpage
   \begin{figure}[h] 
   \centering
   \includegraphics[width=120px]{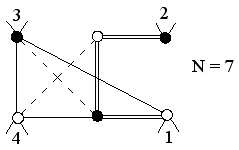} 
  \end{figure}
  \vspace{-25mm}
\begin{equation} \end{equation}
\vspace{5mm}

With an integration by parts to move the numerator factor, and the removal of a double transform, this is simply:

 \begin{figure}[h] 
   \centering
   \includegraphics[width=97px]{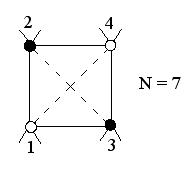} 
  \end{figure}
  \vspace{-25mm}
\begin{equation}\label{eq:n7box1} \end{equation}
\vspace{5mm}

Explicit calculation confirms that this diagram, regarded as a single object in N=7 super-twistor space, gives the correct gravitational 4-amplitude. Note that the super-symmetry plays a maximal r\^{o}le in this  diagram: expanding out the fermionic parts gives an 8-term binomial expansion. (Streamlined methods for performing such integrations are given in section 6 below). This is a strong indication that use of the triangular contour identity has performed  a non-trivial transformation of the N=8 calculus. It is not merely a re-arrangement of terms or a re-writing of super-algebra. We shall examine this further in section 7.

This is essentially the simplest possible twistor diagram that there could be for the amplitude, given that conformal symmetry breaking has to appear. It amounts to integration of   the simple  numerator $I_{\alpha\beta}Z_2^{\alpha}Z_3^{\beta}I^{\gamma\delta}W_{1\gamma}W_{4\delta} = \langle23\rangle[14]$ over a super-volume.
No double poles now arise, the numerator seventh power is natural, and the expression has manifest symmetry in $\{23\}$ and $\{14\}$, which is the best that can be attained within the N=7 formalism.

It should be clear that the N=7 formalism does capture all the information in standard N=8 supergravity. Two N=7 supermultiplets of 128 fields correspond to one N=8 supermultiplet of 256 fields. The identification can be made by  picking out (say) the eighth super-component. We then divide the N=8 supermultiplet of 256 fields  into two classes, those that do not contain an 8 and those that do. These classes naturally map into two dual N=7 128-multiplets. For instance, considering the scalar fields, $\phi^{(8)}_{1234}$ is simply mapped to $\phi^{(7)}_{1234}$, whilst  $\phi^{(8)}_{1238}$ is mapped to the complementary $\phi^{'(7)}_{4567}$ in the dual N=7 multiplet. Thus the 70 N=8 scalar fields are mapped to $35+35$ N=7 scalar fields. For the twistor representation, this means that instead of having a choice between a $Z^{\alpha}$ and a $W_{\alpha}$ representation for a super-field, with the twistor transform mapping between the two equally good representations, either a  $Z^{\alpha}$ or a $W_{\alpha}$ representation is specified, and there is no twistor transform within the N=7 formalism. (But the {\em double} transform is well-defined, and we have already used it in the diagram transformation above.) In particular, the gravitons themselves have to be represented by functions of homogeneity $+2$ in either $Z^{\alpha}$ or $W_{\alpha}$, depending on helicity. That is indeed what the diagram (\ref{eq:n7box1}) specifies.

How have we contrived to squeeze the {\em three} inverse operators in DeWitt's formula (\ref{eq:dewitt}) into a formalism which in gauge theory expresses only {\em two}? This question could have been addressed in a more elementary way. We are given the amplitude with its $(s_{12}s_{13}s_{14})^{-1}$ factors, and seek to encode it in a twistor calculus. If the external fields are represented by functions of homogeneity degree  $(-6)$, analogous to the starting-point for  gauge-field theory with fields of degree $(-4)$, then indeed we are driven into the splitting and double pole solution.  But twistor theory allows another representation of gravitons, by functions of degree $(+2)$. In this representation, the contraction of the spinor indices in $\psi_{1ABCD}\psi_2^{ABCD}$ translates into the operation  $( I_{\alpha\beta}\frac{\partial}{\partial Z_1^{\alpha}}\frac{\partial}{\partial Z_2^{\beta}})^4$. These derivatives in the numerator can cancel the problematic inverse operator $s_{12}^{-1} =  (I_{\gamma\delta}Z_1^{\gamma}Z_2^{\delta})^{-1}(I^{\alpha\beta}\frac{\partial}{\partial Z_1^{\alpha}}\frac{\partial}{\partial Z_2^{\beta}})^{-1}$.  This exploits the subtle asymmetry in the DeWitt expression, in which the 
inverse operator $s_{12}^{-1}$ is not on the same footing as the others. The price paid for this choice of representation is the eight-term summation of terms; but this is packaged into a single term through the magic of N=7 super-symmetry.

This description still does not quite explain why the inverse derivative $s_{12}^{-1}$ can somehow be made innocuous, apparently not actually requiring any integration at all. The reason why this is possible is that a twistor function of degree $(+2)$  {\em already} embodies a great deal of integration! The Penrose transform requires such a function to be differentiated four times to get the physical Weyl spinors such as $\psi_{1ABCD}.$ They are in a sense potentials, but of a kind that never introduces the spurious degrees of freedom that plague the conventional potentials of gauge theory and gravity. Thus this geometric realization of DeWitt's formula actually goes to the the heart of the twistor representation.
 
On the other hand, the N=7 formalism completely fails to embody one of the main ideas of the twistor representation, that the negative and positive helicity gravitons should be represented in the same twistor space. Considerable effort has gone into finding solutions to this `googly' problem, but the formalism here seems to abandon any hope of so doing, with gravitons of different helicities behaving like completely separate entities. We seem  also to have abandoned  one of the main objectives described by Nguyen, Spradlin, Volovich and Wen (2009),  that of representing the full $S_n$ symmetry of gravitational amplitudes. Hopefully, since all the information of N=8 theory is actually preserved in the N=7 formalism,  some way will be found to  bridge from 7 to 8 and  recover the full symmetry. Meanwhile we shall pursue the computational advantages of the new formalism.

\section{Evaluation of supersymmetric diagrams}
We will now study the N=7 diagram and its evaluation, first putting it in a more convenient order, so that
  \begin{figure}[h] 
   \centering
   \includegraphics[width=92px]{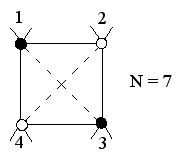} 
  \end{figure}
\vspace{-25mm}
\begin{equation} \label{eq:n7box} \end{equation}
\vspace{10mm}

gives the amplitude $M(1^+2^-3^+4^-).$ How do we evaluate this in an efficient manner?  In principle, every super-boundary has to be expanded as a formal series and then the resulting fermionic numerators combined. In the case of N=4 theory, the diagram 
 \begin{figure}[h] 
   \centering
   \includegraphics[width=91px]{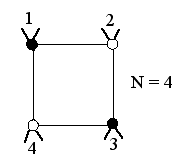} 
  \end{figure}
\vspace{-25mm}
\begin{equation} \label{eq:n4box} \end{equation}
\vspace{10mm}

with fields of classical homogeneity 0 on the outside, expands to a five-term series of  N=0 diagrams
 \begin{figure}[h] 
   \centering
   \includegraphics[width=382px]{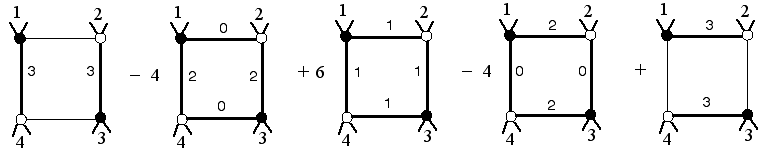} 
  \end{figure}

where the binomial coefficients (and the alternating signs) are the result of formal integration of the fermionic parts of the super-twistors. Likewise, the N=7 diagram expands to an eight-term binomial sum of N=0 diagrams. However, we do not want to do this algebraic computation in detail, and in practice we use the essential property of supersymmetric algebra which matters here --- the property of mimicking the effect of derivative operations. This is most easily shown for N=1.  We have that

 \begin{figure}[h] 
   \centering
   \includegraphics[width=87px]{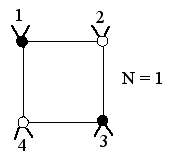} 
  \end{figure}
    \vspace{-25mm}
\begin{equation} \end{equation}
\vspace{5mm}

expands to the two N=0 diagrams (with an essential minus sign from the anticommuting elements):

 \begin{figure}[h] 
   \centering
   \includegraphics[width=165px]{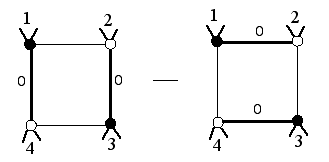} 
  \end{figure}
    \vspace{-25mm}
\begin{equation} \end{equation}
\vspace{5mm}

again assuming that there are no super-components in the exterior fields, i.e.\ that these are all of classical homogeneity $(-3)$. This can be thought of as representing the equivalent of M\"{o}ller scattering for (massless) quarks.  Now we observe that this anti-symmetrized sum is exactly the same as is obtained by the operation $\langle24\rangle/\langle13\rangle$ applied to the all-boundary N=0 diagram, so that the amplitude is
\begin{equation}\frac{\langle24\rangle}{\langle13\rangle} \frac{\langle 13\rangle^4}{\langle12\rangle\langle23\rangle\langle34\rangle\langle41\rangle} =  \frac{\langle 24\rangle \langle13\rangle^3}{\langle12\rangle\langle23\rangle\langle34\rangle\langle41\rangle} \label{eq:moller}\end{equation}
agreeing with the sum of the two separate M\"{o}ller terms: $\langle13\rangle[24]s_{14}^{-1} + \langle13\rangle[24]s_{12}^{-1}$.

 It is worth noting that from this point of view,  the anti-commutation in the formal super-algebra stems from and expresses the anti-symmetry of the infinity twistor $I_{\alpha\beta}$, which is in turn a consequence of the representation of space-time points by lines in projective twistor space. 

For N=4 this occurs four times over, so that the N=4 diagram (\ref{eq:n4box}) is given by $\langle24\rangle^4/\langle13\rangle^4$ times the corresponding N=0 diagram, correctly giving an amplitude
\begin{equation}\frac{\langle24\rangle^4}{\langle13\rangle^4} \frac{\langle 13\rangle^4}{\langle12\rangle\langle23\rangle\langle34\rangle\langle41\rangle} =  \frac{\langle 24\rangle^4}{\langle12\rangle\langle23\rangle\langle34\rangle\langle41\rangle} \, .\end{equation}

For the N=7 diagram (\ref{eq:n7box}) we may likewise simply read off the result (taking into account the two numerator factors $\langle13\rangle[24]$)  as:
\begin{equation}\frac{\langle24\rangle^7}{\langle13\rangle^7} \frac{\langle 13\rangle [24]\,\, \langle13\rangle^4}{\langle12\rangle\langle23\rangle\langle34\rangle\langle41\rangle} =  \frac{\langle 24\rangle^7[24]}{\langle13\rangle^2\langle12\rangle\langle23\rangle\langle34\rangle\langle41\rangle}\, ,\end{equation}
which is correct.

A slightly different evaluation comes from relating the N=7 diagram to the corresponding N=4 diagram, which we can identify with $A(1^+2^-3^+4^-)$, and then considering the numerator factors as arising from applying the operator $s_{12}$. This gives:
\begin{equation}\frac{\langle24\rangle^3}{\langle13\rangle^3 } s_{12} \, \frac{\langle13\rangle^2}{\langle12\rangle\langle34\rangle}\, A(1^+2^-3^+4^-)\, =\,  s_{12}\,A(1^+2^-4^-3^+)\,A(1^+2^-3^+4^-)\end{equation}
which is the KLT form of the gravitational amplitude.

These methods extend to higher amplitudes. For example, the 5-field N=4 gauge-theoretic MHV amplitude $A(1^+2^-3^+4^+5^-)$ is represented by
 
  \begin{figure}[h] 
   \centering
   \includegraphics[width=109px]{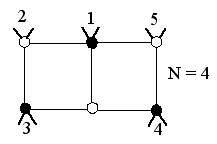} 
  \end{figure}
    \vspace{-25mm}
\begin{equation} \end{equation}
\vspace{5mm}

with the external fields all of classical homogeneity 0. This would, if expanded out in individual terms, require extensive super-algebra and then the summation of a trinomial expression. This is all cut short by observing that 
 the effect of the supersymmetric algebra is exactly the same as applying the operator $(\langle25\rangle/\langle34\rangle)^4$ to the N=0 diagram:

   \begin{figure}[h] 
   \centering
   \includegraphics[width=97px]{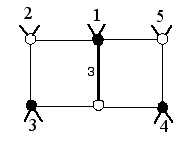} 
  \end{figure}
    \vspace{-25mm}
\begin{equation} \end{equation}
\vspace{5mm}

Thus we may correctly write down in one step:
\begin{equation}A(1^-2^+3^-4^-5^+) = \frac{\langle25\rangle^4}{\langle12\rangle\langle23\rangle\langle34\rangle\langle45\rangle\langle51\rangle}\, .\end{equation}
Such methods lie behind the claim made in (Hodges 2006) that all the eight-gauge-field amplitudes could simply be written down directly from N=4 twistor diagrams. They will be used freely in the ensuing sections.

\newpage

\section{BCFW recursion on N=7 elements}

In this section we shall show that every MHV tree-level interaction can be derived by an N=7 super-symmetric BCFW recursion. 
The recursion is built up from the  three-amplitudes, which we shall now find. To express these as diagrams requires care since there are no twistor transforms within the N=7 formalism; an external field is defined as being either a twistor or a dual twistor (super-)function. But {\em double} transforms are well-defined, and allow us to write the three-amplitudes as:

 \begin{figure}[h] 
   \centering
   \includegraphics[width=81px]{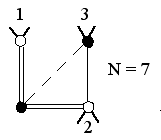} 
  \end{figure}
    \vspace{-25mm}
\begin{equation} \end{equation}
\vspace{5mm}

and its conjugate. Here 1, 2, 3, are super-functions of degree $(+2)$. For gravitons, i.e.\ with bosonic homogeneity $(+2)$ outside and hence all the supersymmetry on the internal lines, this diagram correctly corresponds to amplitude $\langle12\rangle^6\langle23\rangle^{-2}\langle31\rangle^{-2}$.

 This expression can be seen to be symmetric in $\{12\}$ by taking a double twistor transform on the 2 field and moving the numerator factor across using integration by parts; in fact there are several possible forms.  The apparent lack of economy in the representation comes from the fact that the diagrams embody the operation of taking a {\em derivative} of the external fields: thus the amplitude depends on each $f_i(W)$ only through $I_{\alpha\beta}\frac{\partial}{\partial W_{\beta}} f_i(W).$ It is this derivative, for instance, which determines the  `non-linear graviton' of (Penrose 1976). We could write the three-amplitude without any double transforms as  $f_3(Z)[W.Z]_0 f_1(W)Z^{\alpha}I_{\alpha\beta}\frac{\partial}{\partial W_{\beta}} f_2(W)$, and if we invented new notation we could indicate this diagrammatically. There is little point in so doing, however, as in  all higher amplitudes the derivative is naturally taken into the interior of the diagram and the double transform removed.
 
Now we note a remarkable fact. BCFW composition, applied blindly as a rule to these 3-amplitudes, gives the correct four-field amplitude. Pivoting on (23) there is now only {\em one} term for $M(1^+2^-3^+4^-)$, namely $M(1^+2^-k)\circ M(3^+4^-k)$, and that gives
\newpage
 \begin{figure}[h] 
   \centering
   \includegraphics[width=261px]{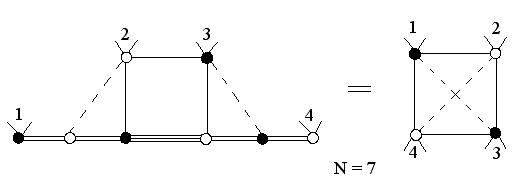} 
  \end{figure}
  \vspace{-25mm}
\begin{equation} \end{equation}
\vspace{5mm}

as claimed.

Before progressing to study further amplitudes, we remark on the significance of this agreement.
The four gravitons  are being considered as the seven-fold super-symmetrized
partners of spin-$\frac{3}{2}$ fields of the opposite helicity type. From this point of view, the four-graviton interaction is being treated as the super-symmetrized version of a process in which four spin-$\frac{3}{2}$ fields exchange gravitons. The amplitude for {\em this} process comes from treating (\ref{eq:n7box}) as an N=0 diagram, i.e.\ with four twistor functions of degree $(-5)$ on the outside, and is:
\begin{equation}\frac{\langle13\rangle^5 [24]}{\langle 12\rangle\langle 23\rangle\langle 34\rangle\langle 41\rangle}\, .\label{eq:gravitinos}\end{equation}

It is worth checking that this agrees with the calculation from N=8 theory. Putting the same external functions into the general formula (\ref{eq:n8form})  we obtain two terms which will indeed sum to the same result. We can look in more detail at how this agreement occurs, by expanding out just the eighth super-component and comparing with the N=7 expressions. We find that the first of the two N=8 terms gives rise to just two N=7 terms, of which one is the same as the N=7 diagram, and the other cancels the second of the two N=8 terms. This cancellation occurs through the square symmetry identity (\ref{eq:twoterm}), which is just a disguised form of the triangular contour identity.

This is again a signal that the agreement of the N=7 diagram recursion with the N=8 calculation is non-trivial. To investigate this agreement further it is useful to compare it with the analogous structure in gauge theory. We can follow all the steps above, writing down N=3 three-amplitudes, and applying BCFW recursion to them. But the results of this procedure are simply false; indeed it fails at the very simplest example, the (massless) quark-gluon analogue of M\"{o}ller scattering, which is the gauge-theory analogue of the process (\ref{eq:gravitinos}) above. Recursion based on the N=3 three-amplitudes  captures just one of the two diagrams in (\ref{eq:moller}), and not the anti-symmetrized sum.

Thus the N=4 theory has terms which do {\em not} cancel in the passage to N=3, confirming that the cancellation of terms in the passage from N=8 to N=7 is no triviality, but depends on something that makes gravity essentially different from gauge theory.

This explicit comparison also suggests what that essential difference might be. We may study the behaviour of the momentum-shifted amplitude as a function of the complex parameter $z$, i.e.\ for $|2\rangle|2] \rightarrow |2\rangle|2] + z|3\rangle|2], \,\,\,\,|3\rangle|3] \rightarrow |3\rangle|3] - z|3\rangle|2]$.  The massless quark-gluon M\"{o}ller amplitude (\ref{eq:moller}) fails to satisfy the condition $\lim_{z\rightarrow \infty} A(z) =0$. Hence the partial-fraction argument which justifies BCFW is not valid. In contrast, the amplitude (\ref{eq:gravitinos})  for the gravitational analogue does satisfy this condition.\footnote{In this Version 2, a new section 13 shows how this observation does indeed extend to showing  the validity of the N=7 recursion for all tree amplitudes. The author is most grateful to Henriette Elvang for this valuable contribution.}

This simple power-counting difference may be connected with the fact that the complete amplitude for N=8 gravity  behaves better than is required for the validity of BCFW recursion, decreasing asymptotically like $z^{-2}$. As is well known, this implies the existence of extra relations.
As discussed in (Nguyen, Spradlin, Volovich, Wen 2009), one of the most desirable features of a recursive procedure would be that it conforms to this $z^{-2}$ behaviour at each stage, which can be taken as a sign that the procedure has made use of these extra relations. It would be interesting to know if the N=7 recursion calculus is in fact an effective way of ensuring this conformity. We leave this as an open question. 

In the next section we shall check by less direct means that the agreement of our new N=7 recursion continues to agree with the standard N=8 recursion, at least for all MHV processes. We shall not look at the detail of the cancellations, but simply show that the N=7 recursion calculus gives agreement with known results.

\section{All MHV amplitudes by N=7 BCFW}

 We have already checked that applying the BCFW rule to these 3-amplitudes generates the 4-field amplitude.
 We shall now continue by considering the  5-graviton amplitude.   
 Application of standard N=8 BCFW results in a sum of {\em three} terms, each replete with double poles, all of which cancel, but only through the application of Schouten identities both in the $\langle \, \rangle$ and the $[\,]$ spinors. Of course, this cancellation can be facilitated by noting that double poles cancel in the 4-field sub-amplitudes, and using this fact recursively. The calculation in (Cachazo and Svr\u{c}ek 2005) does this.

 When thus simplified, the amplitude can be expressed in BGK form as a two-term sum:
\begin{equation}M(1^+2^- 3^+ 4^+ 5^-) =\langle 25\rangle^7 \frac{[45]\langle35\rangle[23]\langle42\rangle - [35]\langle45\rangle[42]\langle23\rangle}{\langle 12\rangle \langle 13\rangle \langle14\rangle \langle 15\rangle \langle 23\rangle \langle 24\rangle  \langle 34\rangle \langle 35\rangle \langle 45\rangle}\, .\label{eq:bgk5}\end{equation}
We will now compare this procedure with the result of applying the  N=7 BCFW recursion, using 5 and 1 as pivot fields. There are only two terms, arising from
$M(4^+5^-k)\circ M(1^+2^-3^+k)$ and $M(3^+5^-k)\circ M(1^+2^-4^+k)$. The four-field sub-amplitudes need no simplification before being substituted in, for they are  already natural single-term expressions. The diagrammatic form of the BCFW joining process then yields the complete amplitude as the sum of two diagrams:
 \begin{figure}[h] 
   \centering
   \includegraphics[width=226px]{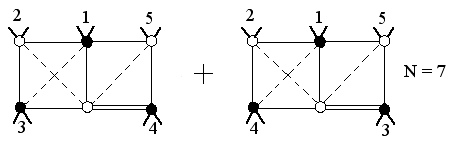} 
  \end{figure}
    \vspace{-25mm}
\begin{equation}\label{eq:n7box5} \end{equation}
\vspace{5mm}

Remarkably, these diagrams correspond directly to the two terms in the BGK formula. To establish this, consider the first diagram (the second is the same with $3\leftrightarrow 4$). We may first show that
the N=0 diagram
 \begin{figure}[h] 
   \centering
   \includegraphics[width=92px]{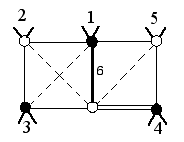} 
  \end{figure}
    \vspace{-25mm}
\begin{equation} \end{equation}
\vspace{5mm}

 corresponds to  \begin{equation} \frac{\langle 34\rangle^6[23][45]}{\langle23\rangle  \langle 12\rangle \langle 13\rangle \langle 14\rangle\langle 15\rangle \langle 45\rangle}\, ,\end{equation}
which is straightforward, Then we use the expansion principle of N=7 diagrams, described in section 7, and so multiply this by $(\langle25\rangle/\langle34\rangle)^7$. The result follows.

If we relate the N=7 diagrams (\ref{eq:n7box5}) to the corresponding N=4 diagrams, we find that the two diagrams correspond directly to the two terms in the KLT formula:
\begin{eqnarray}M(1^+ 2^- 3^+ 4^+ 5^-) &=&s_{23} s_{45}A(1^+ 2^- 3^+ 4^+ 5^-)A(1^+4^+5^-2^-3^+)  \nonumber \\
&+&s_{24} s_{35}A(1^+ 2^- 4^+ 3^+ 5^-)A(1^+3^+5^-2^-4^+) \, .\end{eqnarray}

None of the above actually depended on the use of the twistor diagrams, and could have been written out in standard terms using shifted (super-)momenta. However, one feature of the twistor diagram expressions which cannot be seen in standard terms is the structure of {\em pure numerators} which capture the effect of the gravitational interaction.  Even more strikingly, the twistor diagrams (\ref{eq:n7box5}) can naturally be {\em added} to give a total geometric object analogously to the discussion of the five-field amplitude in gauge theory in section 4. We can use this fact to show an  emergent feature of the N=7 formalism: this 5-graviton amplitude, regarded as a single total object, actually has $S_3 \times S_2$ symmetry structure, being symmetric in $\{134\}$ and in $\{25\}$.

 The argument is similar to the analogous result for gauge theory at (\ref{eq:5sym}), but now that twistor transforms do not exist we must go back to the more fundamental triangular contour identity. In fact we need two rules: the first follows from allowing integration by parts (equivalent to the addition of a total derivative to the integrand), which means that we can always move numerator factors thus:
  \begin{figure}[h] 
   \centering
   \includegraphics[width=84px]{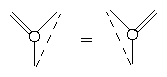} 
  \end{figure}
    \vspace{-15mm}
\begin{equation} \end{equation}

 The second rule is the triangular contour identity, which we can write as: 
   \begin{figure}[h] 
\centering
   \includegraphics[width=132px]{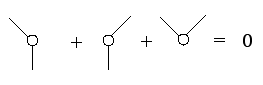} 
  \end{figure}
  \vspace{-20mm}
\begin{equation} \end{equation}

These two rules can be used, rather as in graph theory or knot theory, to define classes of equivalent diagram sums.

 In the case of the 5-graviton amplitude, with diagrammatic representation (\ref{eq:n7box5}),  adding and subtracting the diagram
 
\begin{figure}[h] 
   \centering
   \includegraphics[width=123px]{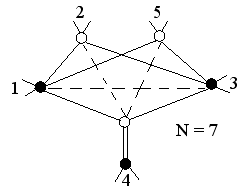} 
  \end{figure}
  \vspace{-25mm}
\begin{equation} \end{equation}
\vspace{5mm}

has the effect of transposing 1 and 3. 
It is easy to see that this simply corresponds to adding and subtracting the expression
\begin{equation}\frac{[24][54]\langle25\rangle^7}{\langle12\rangle\langle13\rangle\langle15\rangle\langle23\rangle\langle35\rangle\langle14\rangle\langle34\rangle}\end{equation}
from (\ref{eq:bgk5}),  but the effect of the diagram formalism is to make this algebraic step into  something  geometrical. 

The 6-graviton MHV amplitude $M(1^+ 2^- 3^+ 4^+ 5^+6^-)$ gives a straightforward continuation of these features. Using $(6^-1^+)$ as pivot, there are only three terms in the N=7 BCFW recursion, namely \begin{eqnarray}M(6^-3^+k)\circ M(1^+2^-4^+5^+k) &+& M(6^-4^+k)\circ M(1^+2^-5^+3^+k)\nonumber\\ &+& M(6^-5^+k)\circ M(1^+2^-3^+4^+k)\, ,\end{eqnarray} and each of the 5-field sub-amplitudes only needs two terms. Thus the whole amplitude may be written as the sum of six diagrams: 
 \begin{figure}[htbp] 
   \centering
   \includegraphics[width=152px]{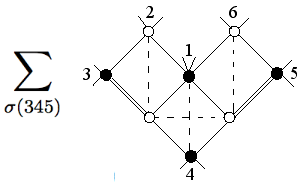} 
  \end{figure}
  \vspace{-25mm}
\begin{equation} \end{equation}
\vspace{5mm}

This  is directly equivalent to
\begin{equation}\sum_{\sigma(345)} \frac{\langle26\rangle^7 [23][56][4|5+6|1\rangle}{\langle23\rangle\langle34\rangle\langle45\rangle\langle56\rangle\langle12\rangle\langle13\rangle\langle14\rangle\langle15\rangle\langle16\rangle}\, .\label{eq:bgk6}\end{equation}
This  is exactly the BGK expression, as restated and proved by Mason and Skinner (2009). 
Its key feature is the factor $[4|5+6|1\rangle = - [4|2+3|1\rangle$, which corresponds directly to the completely internal numerator factor in the centre of the diagram.

Again, by relating the N=7 diagrams to analogous N=4 diagrams, and expressing the numerators in terms of $s_{ij}$ operators, it is straightforward to re-express this six-term result as the KLT formula. It would be of interest to see whether this correspondence could be interpreted in terms of string vertex operators, thus showing some connection between the twistor diagrams and  closed string structure. 

It is also the case that this sum, with its manifest  $S_3$ symmetry, is actually a geometrical object with $S_4$ symmetry, provided we are allowed to use the graphical transformation rules. The explicit proof of symmetry in
$\{1345\}$ is omitted here, as it does not show any particularly new or interesting feature. Clearly the more important next step is  to demonstrate that the symmetry $S_{n-2}$ holds in general --- and hopefully also to learn from such a proof something more about the nature of the geometrical object with this symmetry.

It should be clear how the diagrams generalise to all $n$. A chain of numerator factors develops, corresponding precisely to the numerator factors in the BGK formula.

 A diagram for the case $n=8$ is given here; the complete amplitude is the sum of the 120 permutations of this over $\{34567\}$.

 \begin{figure}[htbp] 
   \centering
   \includegraphics[width=135px]{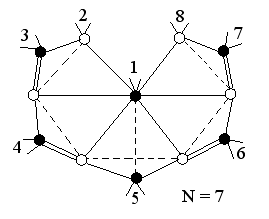} 
  \end{figure}
  \vspace{-25mm}
\begin{equation} \end{equation}
\vspace{5mm}

To summarize: agreement with known results proves that the N=7 recursion is correct at least for MHV (and indeed anti-MHV) gravitational amplitudes.

Lastly we will examine the asymptotic behaviour in the shift parameter $z$. The shift we are interested in is the shift that would be used with non-supersymmetric BCFW pivoted on $\{6^-1^+\}$.  This shift  is $$6] \rightarrow 6] + z\, 1]; \quad 1\rangle \rightarrow 1\rangle - z\,6\rangle\, .$$ Note that this is the dual of the shift that is used in the N=7 BCFW recursion. 
Under this shift the expression (\ref{eq:bgk6}) for the 6-graviton MHV amplitude becomes: 
\begin{eqnarray}&&\sum_{\sigma(345)} \frac{\langle26\rangle^7 [23]\{[56] + z[51]\}\{[4|5+6|1\rangle - z [4|1+5|6\rangle\}}{\begin{array}{l}\langle23\rangle\langle34\rangle\langle45\rangle\langle56\rangle\langle16\rangle\\ \quad \{\langle12\rangle-z\langle62\rangle\}\{\langle13\rangle-z\langle63\rangle\}\{\langle14\rangle-z\langle64\rangle\}\{\langle15\rangle-z\langle65\rangle\}\end{array}}\, \nonumber\\ && \end{eqnarray}
So every one of the six terms behaves as $z\rightarrow\infty$ like $z^2/z^4 = z^{-2}$. For larger $n$ the pattern is the same, with one more numerator power and one more denominator power for each increase in $n$. Thus for MHV amplitudes, an aspect of the N=7 formalism is that the $z^{-2}$ behaviour of the complete amplitude is manifested in every term arising in the recursion.

\newpage

\section{A simple recursion  for  MHV amplitudes}

In this section we shall cast these results into a more conventional momentum-space form. We shall also step outside the pure N=7 theory, to borrow the insight from the N=8 theory that
the MHV amplitudes are naturally written in terms of  helicity-independent reduced amplitudes $\bar{M}_n$:
\begin{equation}M_n(1^-2^-3^+4^+\ldots n^+) =\langle12\rangle^{8}\bar{M}_n(1234\ldots n) \end{equation}
so that $\bar{M}_n$ is totally symmetric in $n$ arguments.

Thus $\bar{M}_3(123) = ( \langle12\rangle\langle23\rangle\langle31\rangle)^{-2}$,
$\bar{M}_4= [34](\langle12\rangle^2 \langle23\rangle\langle34\rangle\langle41\rangle\langle13\rangle\langle24\rangle)^{-1},$ and
\begin{equation}\bar{M}_5(12345) = \frac{[34]\langle24\rangle[12]\langle31\rangle - [24]\langle34\rangle[31]\langle12\rangle}{\langle 12\rangle \langle 13\rangle \langle 14\rangle \langle 15\rangle \langle 23\rangle \langle 24\rangle \langle 25\rangle \langle 34\rangle \langle 35\rangle \langle 45\rangle}\, .\label{eq:mbar5}\end{equation}

Now we can write out in spinor terms the effect of using  the N=7 BCFW formalism, as a simple recursive formula. Our model is the gauge-theoretic analogue, where the recursive formula
\begin{equation}A(123\ldots n-1, n) = \frac{\langle n-1,1\rangle}{\langle n-1,n\rangle\langle n1\rangle}A(123\ldots n-1)\end{equation}
immediately produces the Parke-Taylor formula as solution. The N=7 recursion, when written in terms of the $\bar{M}_n$, is:
\begin{equation}
\bar{M}_n(123\ldots n-1,n) = \sum_{p=3}^{n-1} \frac{[pn]}{\langle p n\rangle}\frac{\langle1p\rangle\langle 2p\rangle}{\langle1n\rangle\langle 2n\rangle}\bar{M}_{n-1}(\hat{1}_{(p)} 23\ldots \hat{p}\ldots n-1)\, ,\label{eq:recursion}\end{equation}
where \begin{equation}\hat{1}_{(p)}] =  \frac{ (1+n)|p\rangle}{ \langle1p\rangle}, \quad\quad \hat{1}_{(p)}\rangle = 1\rangle, \quad\quad \hat{p}] = \frac{(p + n)|1\rangle}{ \langle p1\rangle}, \quad\quad \hat{p}\rangle =p\rangle \, ,\end{equation}
so that $\hat{1}_{(p)} + \hat{p} = 1 + p + n$.
The notation $\hat{1}_{(p)}$ is used to emphasise that the shifted momentum $\hat{1}$ is different in each of the $(n-3)$ terms, depending on $p$. 

This relation appears to differ from others in the literature which are similar in form; thus the formula of (Bedford, Brandhuber, Spence and Travaligni 2005) leads to sums over $(n-2)!$ terms; those from  conventional BCFW have double poles; the relation in (Nguyen, Spradlin, Volovich and Wen 2009) also gives rise to double poles, though it is of particular interest for its derivation from  physical ideas.
 
But no claim is made here that this recursion relation is new, since it is the natural relation satisfied by the BGK formula and may well have been observed before. All that is new is that it here been derived from a principle, that of N=7 recursion.

\section{A new form for the MHV amplitude}

This section does not use the N=7 formalism directly, but is strongly influenced by what it has brought to light. It is also influenced by remarks of Marcus Spradlin, Anastasia Volovich, Nima Arkani-Hamed and others who have emphasised the significance of the inverse soft factors which build up the $n$-graviton MHV amplitudes. In the diagrams we see the addition of triangles in going from $n$ to $n+1$ gravitons, and these triangles are equivalent to such factors. The diagrams can be thought of as the result of adding such triangles in all possible ways to the original 3-amplitude. 

We are also mindful of the goal, emphasised by these authors, of representing the complete $S_n$ symmetry of the amplitudes. By finding a structure with $S_{n-2}$ symmetry but requiring only $(n-3)!$ summands to represent it, we have taken a small step in this direction. As this structure is strongly geometrical, residing in the homology of certain many-dimensional integrals, we are led to look for emergent geometrical structure in the amplitudes $\bar{M}_n$.

Certainly the complete symmetry of $\bar{M}_5$, as given at (\ref{eq:mbar5}), can be given a geometrical characterization. The denominator is manifestly completely antisymmetric, and the numerator has  complete anti-symmetry since it is just $\epsilon_{abcd}\,p_1^a\,p_2^b\,p_3^c\, p_4^d$, where $\sum_{i=1}^5 p_i=0$.  For higher $n$, the BGK formula does not seem to suggest any analogous geometrical pattern. But a striking feature of the recursive formula we have distilled from the N=7 calculus is that the $[ij]$ spinors only enter in the combination $[ij]/\langle ij\rangle$, which appears as a multiplicative factor.  This is a momentum {\em phase factor}, since for real momenta, it is a complex number with modulus 1. This suggests that there may be other solutions to the recursive relation in which this phase factor plays a leading r\^{o}le, and which will point to new geometrical pictures based on the physical concept of the inverse soft factor.

For this purpose we make a new definition:  
\begin{equation}\psi^i_j = \frac{[ij]}{\langle ij\rangle} \,\,({\hbox{for  }} i \ne j)\, , \quad \psi^i_i =0.
\label{eq:psidefn}\end{equation}
 So $\psi^i_j = \psi^j_i$. The use of upper and lower indices has no significance except for typographical convenience when expressing antisymmetrization. {\em Note:} Such anti-symmetrization, when indicated below by the use of square brackets round $n$ indices, is not accompanied by a $1/n!$ factor; thus $\psi^1_{[4}\psi^2_{5]}$ means $\psi^1_4\psi^2_5 - \psi^1_5\psi^2_4$.

This gives yet another way of writing the existing results: 
\begin{eqnarray}\bar{M}_3(123) &=& \frac{1}{\langle12\rangle \langle23\rangle\langle31\rangle\,\, \langle12\rangle \langle23\rangle\langle31\rangle}\,,\\
\bar{M}_4(1234) &=& \frac{\psi^1_4}{\langle12\rangle \langle23\rangle\langle31 \rangle\,\, \langle23\rangle\langle34\rangle\langle42\rangle}\,,\end{eqnarray}
which is actually completely symmetric in $\{1234\}$, and
\begin{equation}\bar{M}_5(12345) = \frac{\psi^1_{[4}\psi^2_{5]}}{\langle12\rangle \langle23\rangle\langle31 \rangle\,\,\langle34\rangle\langle45\rangle\langle53\rangle}\,,\end{equation}
which is actually completely symmetric in $\{12345\}$. Note that the symmetry of $\bar{M}_4(1234)$ and $\bar{M}_5(12345)$ depends upon the momentum conservation condition holding. 
Before going on to $\bar{M}_6$ we shall state and prove:

{\bf Lemma (6-point double spinor identity)}: Given six spinors and six dual spinors, denoted by $|i\rangle, i]$ in the usual way, subject only to the condition that $\langle ij\rangle \ne 0$ when $i\ne j$,
\begin{equation}\frac{\psi^1_{[4}\psi^2_{5}\psi^3_{6]}}{\langle12\rangle \langle23\rangle\langle31\rangle\,\,\langle45\rangle\langle56\rangle\langle64\rangle}\label{eq:doublespinor}\end{equation}
is completely symmetric in $\{123456\}$.

{\bf Proof:} The given expression clearly has the symmetry of the $K(3,3)$ graph, which is a subgroup of order 72 of the full permutation group $S_6$.  To show the complete symmetry we need only show invariance under (e.g.) $2\leftrightarrow 5$, and this can be done as follows. Consider the difference
\begin{equation} \frac{\psi^1_{[2}\psi^3_4\psi^5_{6]}}{\langle13\rangle \langle35\rangle\langle51\rangle\langle24\rangle\langle46\rangle\langle62\rangle} - \frac{\psi^1_{[4}\psi^2_5\psi^3_{6]}}{\langle12\rangle \langle23\rangle\langle31\rangle\langle45\rangle\langle56\rangle\langle64\rangle}\, .\end{equation}
We want to show this vanishes, which is equivalent to showing that
$$\langle12\rangle\langle45\rangle\langle56\rangle\langle23\rangle \psi^1_{[2}\psi^3_4\psi^5_{6]} - \langle15\rangle\langle35\rangle\langle24\rangle\langle26\rangle \psi^1_{[4}\psi^2_5\psi^3_{6]} =0\, .$$
In the first term, write the various products such as $[12][34][56]$ in terms of those products which appear in the second term, for instance by expanding  $[12][34][56]$  as $([15][26]-[16][25])[34]$. We then collect together similar terms. There are two types of term, depending on whether or not they contain $[25]$. 

Consider first the terms which are multiples of $[15][26][34]$. There are two of these, one  from expanding $[12][34][56]$, and one from $[15][26][34]$ itself. These give rise to
$$\left\{\frac{\langle45\rangle\langle23\rangle}{\langle34\rangle} - \frac{\langle35\rangle\langle24\rangle}{\langle34\rangle}\right\}[15][26][34]= \langle25\rangle [15][26][34] \,.$$
Now consider the terms which are multiples of $[16][25][34]$. There are four of these, giving rise to
$$\left\{ \frac{\langle12\rangle\langle56\rangle}{\langle16\rangle}- \frac{\langle45\rangle\langle23\rangle}{\langle34\rangle} + \frac{\langle12\rangle\langle56\rangle\langle45\rangle\langle23\rangle}{\langle16\rangle\langle25\rangle\langle34\rangle}- \frac{\langle15\rangle\langle26\rangle\langle35\rangle\langle24\rangle}{\langle16\rangle\langle25\rangle\langle34\rangle}\right\}[16][25][34]$$
$$= \left\{\frac{\langle12\rangle\langle56\rangle}{\langle16\rangle}-\frac{\langle15\rangle\langle26\rangle\langle45\rangle\langle23\rangle}{\langle16\rangle\langle25\rangle\langle34\rangle}- \frac{\langle15\rangle\langle26\rangle\langle35\rangle\langle24\rangle}{\langle16\rangle\langle25\rangle\langle34\rangle} \right\}[16][25][34]$$
$$= \left\{\frac{\langle12\rangle\langle56\rangle}{\langle16\rangle}- \frac{\langle15\rangle\langle26\rangle}{\langle 16\rangle}\right\} [16][25][34] = \langle25\rangle [16][25][34] $$
The other terms are obtained by $1\leftrightarrow 3, 4\leftrightarrow 6$, and the difference is thus
$$\langle25\rangle([14][25][36]-[16][25][34]+[14][26][35]-[16][24][35]+[15][26][34]- [15][24][36]) = 0$$
Note that the calculus uses nothing but three-term Schouten identities in both kinds of spinors, and that it is an algebraic identity valid irrespective of whether the six null vectors $|i\rangle i]$ sum to zero.

This completely symmetrical quantity is in fact the six-field MHV amplitude:

{\bf Theorem (6-graviton MHV amplitude formula):}
\begin{equation}\bar{M}_6(123456) =  \frac{\psi^1_{[4}\psi^2_5\psi^3_{6]}}{\langle12\rangle \langle23\rangle\langle31\rangle\,\,\langle45\rangle\langle56\rangle\langle64\rangle}\, .\end{equation}
{\bf Proof:} From the recursive formula (\ref{eq:recursion}), 
$$\bar{M}_6(123456) = \frac{[56]\langle15\rangle\langle25\rangle}{\langle56\rangle\langle16\rangle\langle26\rangle} \bar{M}_5(\hat{1}_{(5)} 234\hat{5}) + \hbox{ two similar terms.}$$
If we write 
$$ \bar{M}_5(\hat{1}_{(5)} 234\hat{5}) = \frac{\psi^2_{[3}\psi^{\hat{5}}_{4]}}{\langle13\rangle \langle34\rangle \langle41\rangle \langle12\rangle \langle26\rangle \langle61\rangle }$$
then, on using $\hat{5}] = (5+6)| 1\rangle \langle51\rangle^{-1}$, we obtain the BGK formula (\ref{eq:bgk6}).

On the other hand, we are free to write
$$ \bar{M}_5(\hat{1}_{(5)}234\hat{5}) = \frac{\psi^{\hat{1}_{(5)}}_{[3}\psi^{2}_{4]}}{\langle12\rangle \langle25\rangle \langle51\rangle \langle34\rangle \langle45\rangle \langle53\rangle }$$
and then we have
$$\bar{M}_6(123456)= \frac{\psi^{\hat{1}_{(5)}}_{[3}\psi^{2}_{4]}\psi^6_{5}}{\langle12\rangle \langle26\rangle \langle61\rangle \langle34\rangle \langle45\rangle \langle53\rangle  } + \hbox{ two similar terms.}$$
Now, $\hat{1}_{(p)}] = 1] + 6] \langle6p\rangle\langle1p\rangle^{-1}$. The $1]$ part immediately gives the required formula. Further terms come from $6] \langle6p\rangle\langle1p\rangle^{-1}$. There are six such terms, each containing a factor of form $[6i][6j]$, where $i$ and $j$ are two distinct elements of $\{345\}$. The terms including $[63][65]$ are:
$$\left(\frac{[63]\langle65\rangle\langle15\rangle^{-1}} {\langle13\rangle} \frac{[24][65]}{\langle 24\rangle\langle65\rangle} - \frac{[65]\langle63\rangle\langle13\rangle^{-1}} {\langle15\rangle} \frac{[24][63]}{\langle 24\rangle\langle63\rangle} \right)\frac{1}{\langle12\rangle \langle26\rangle \langle61\rangle \langle34\rangle \langle45\rangle \langle53\rangle}$$
from $p=5$ and $p=3$ respectively, and these cancel. All the others are similar, hence the result.

The 6-point double spinor identity (\ref{eq:doublespinor}) now gives rise to a further result relevant to the amplitude for $n=7$.

{\bf Corollary (7-point double spinor identity):}
Given seven spinors and seven dual spinors, denoted by $|i\rangle, i]$ in the usual way, subject only to the condition that $\langle ij\rangle \ne 0$ when $i\ne j$,
\begin{equation}\frac{\psi^7_{[3}\psi^1_{4}\psi^2_{5}\psi^6_{7]}}{\langle12\rangle \langle26\rangle\langle61\rangle\,\,\langle34\rangle\langle45\rangle\langle53\rangle}\label{eq:7identity}\end{equation}
is completely symmetric in $\{123456\}$.

{\bf Proof:} Expand the numerator as
$$\psi^7_{3}\psi^1_{[4}\psi^2_{5}\psi^6_{7]} + \psi^7_{4}\psi^1_{[5}\psi^2_{3}\psi^6_{7]}+ \psi^7_{5}\psi^1_{[3}\psi^2_{4}\psi^6_{7]}$$ and apply the 6-point double spinor identity (\ref{eq:doublespinor}) to each term, giving
\begin{equation}\frac{\psi^7_{3}\psi^1_{[4}\psi^2_{5}\psi^7_{6]}\langle74\rangle\langle75\rangle\langle36\rangle + \psi^7_{4}\psi^1_{[5}\psi^2_{3}\psi^7_{6]}\langle75\rangle\langle73\rangle\langle46\rangle+ \psi^7_{5}\psi^1_{[3}\psi^2_{4}\psi^7_{6]}\langle73\rangle\langle74\rangle\langle56\rangle}
{\langle12\rangle \langle27\rangle\langle71\rangle\,\,\langle34\rangle\langle45\rangle\langle53\rangle\langle64\rangle\langle65\rangle\langle63\rangle}\, .\label{eq:3456} \end{equation}
The denominator is antisymmetric in $\{3456\}$. The numerator is manifestly antisymmetric in $\{345\}$, and to show its antisymmetry in $\{3456\}$ we need only check antisymmetry in $\{46\}$. Symmetrizing the numerator over $\{46\}$ yields
$$\{\psi^7_{3}\psi^1_{[4}\psi^2_{5}\psi^7_{6]}
+\psi^7_{4}\psi^1_{[5}\psi^2_{3}\psi^7_{6]} + \psi^7_{6}\psi^1_{[5}\psi^2_{3}\psi^7_{4]}+\psi^7_{5}\psi^1_{[3}\psi^2_{4}\psi^7_{6]}\}\langle75\rangle\langle73\rangle\langle46\rangle$$
$$= \psi^7_{[3}\psi^1_{4}\psi^2_{5}\psi^7_{6]}\langle75\rangle\langle73\rangle\langle46\rangle = 0\,.$$
Hence the expression (\ref{eq:3456}) is indeed symmetric in $\{3456\}$. Similarly the expression (\ref{eq:3456}) is symmetric under $3\leftrightarrow1, 4\leftrightarrow1$ etc.\ and so in all of $\{123456\}$.

Note that again the identity does not depend upon momentum conservation, and uses nothing but 3-term Schouten identities.

By the same Schouten identities we also have from (\ref{eq:3456}) that
\begin{equation}\frac{\psi^7_{[3}\psi^1_{4}\psi^2_{5}\psi^6_{7]}}{\langle12\rangle \langle26\rangle\langle61\rangle\,\,\langle34\rangle\langle45\rangle\langle53\rangle} = \sum_{\sigma'(3456)}\frac{\psi^7_3\psi^7_4\psi^1_5\psi^2_6\langle75\rangle\langle76\rangle\langle34\rangle}{\langle12\rangle \langle27\rangle\langle71\rangle\,\,\langle34\rangle\langle45\rangle\langle53\rangle\langle64\rangle\langle65\rangle\langle63\rangle}\label{eq:34562} \end{equation}
where the summation is over the 12 permutations of $\{3456\}$ which give distinct terms.

{\bf Theorem (a 7-graviton MHV amplitude formula):} 
Define the function $R_7(123456;7)$ by
\begin{equation}R_7(123456;7) = \frac{\psi^7_{[3}\psi^1_{4}\psi^2_{5}\psi^6_{7]}}{\langle12\rangle \langle26\rangle\langle61\rangle\,\,\langle34\rangle\langle45\rangle\langle53\rangle}\label{eq:7amplitude}\end{equation}
so that by the preceding result (\ref{eq:7identity}), 
$R_7(123456;7)$ is totally symmetric in $\{123456\}$.
Then $$\bar{M}_7(1234567) = \frac{1}{3}\sum_{i=1}^7R(i,i+1,i+2,i+3,i+4,i+5;i+6)$$
so that given the symmetry of $R$, $\bar{M}_7$ is manifestly totally symmetric.

{\bf Proof:} From the recursive formula (\ref{eq:recursion}),
$$\bar{M}_7(1234567) = \frac{[67]\langle16\rangle\langle26\rangle}{\langle67\rangle\langle17\rangle\langle27\rangle}\bar{M}_6(\hat{1}_{(6)}2345\hat{6}) + \hbox{ 3 similar terms } $$
By choosing a judicious representation of $\bar{M}_6$, this is immediately
\begin{equation}\frac{\psi^{\hat{6}}_{[3}\psi^{\hat{1}_{(6)}}_4\psi^2_{5]}\psi^7_6}{\langle12\rangle\langle27\rangle\langle71\rangle \,\,\langle34\rangle \langle45\rangle\langle54\rangle} + \hbox{ 3 similar terms }  \label{eq:expression}\end{equation}
If the leading parts of $\hat{1}$ and $\hat{p}$ are taken, we obtain 
\begin{equation}\frac{\psi^6_{[3}\psi^{1}_{4}\psi^2_{5]}\psi^7_{6}}{\langle12\rangle\langle27\rangle\langle71\rangle \,\,\langle34\rangle \langle45\rangle\langle54\rangle} + \hbox{ 3 terms obtained by } 6\leftrightarrow  3, 4, 5. \end{equation}
Now we investigate the corrections arising from the shift terms.
First, the part where the shift corrections in {\em both} $\hat{1}$ and $\hat{p}$ are taken in the expression (\ref{eq:expression}) will indeed vanish in the sum. The terms arising from these parts are characterized by containing products like $[7i][7j][7k]$, where $i,j,k$ are distinct members of $\{3456\}$. Thus the terms containing $[73][74][75]$ will arise in the $p=3, p=4,$ and $ p=5$ parts of the sum. The contribution from $p=3$ is:
\begin{eqnarray}&&\frac{1}{\langle12\rangle\langle27\rangle\langle71\rangle\langle45\rangle\langle56\rangle\langle64\rangle}
\left( \frac{[74]\langle17\rangle}{\langle34\rangle\langle13\rangle}\frac{[75]\langle37\rangle}{\langle15\rangle\langle13\rangle} - \{4\leftrightarrow5\}\right) \frac{[26][73]}{\langle26\rangle\langle73\rangle}\nonumber \\
&=& \frac{[73][74][75][26]}{\langle12\rangle\langle27\rangle\langle26\rangle\langle45\rangle\langle56\rangle\langle64\rangle}\frac{1}{\langle13\rangle^2}\left(\frac{1}{\langle34\rangle\langle15\rangle} - \frac{1}{\langle35\rangle\langle14\rangle}\right)\nonumber \\
&=& \frac{[73][74][75][26]}{\langle12\rangle\langle27\rangle\langle26\rangle\langle56\rangle\langle46\rangle\langle13\rangle\langle14\rangle\langle15\rangle\langle34\rangle\langle35\rangle}\nonumber \\
&=& \frac{[73][74][75][26]\quad  \langle36\rangle\langle45\rangle}{\langle12\rangle\langle27\rangle\langle26\rangle\langle13\rangle\langle14\rangle\langle15\rangle\langle36\rangle\langle46\rangle\langle56\rangle\langle34\rangle\langle45\rangle\langle53\rangle}\end{eqnarray}
The denominator in this expression is antisymmetric in $\{345\}$. Hence symmetrizing over $\{345\}$, to account for adding in the $p=4, p=5$ terms, has the effect of antisymmetrizing over $\langle36\rangle\langle45\rangle$, which is zero by the Schouten identity.

Thus we are left with the terms arising from just {\em one} shift correction. Consider those terms which contain $[73][74][26]$ in the expression  (\ref{eq:expression}). These arise from $p=3$ and $p=4$. From $p=3$ we obtain
$$\frac{1}{\langle12\rangle\langle27\rangle\langle71\rangle\langle45\rangle\langle56\rangle\langle64\rangle}\frac{[73]}{\langle73\rangle}\left(\frac{[74]\langle17\rangle[15][26]}{\langle34\rangle\langle13\rangle\langle15\rangle\langle26\rangle} - \{1\leftrightarrow 3\}\right)$$
and from $p=4$ the same with $3\leftrightarrow4$.
Of these terms, some contain $[15]$, some $[35]$ and some $[45]$. The terms containing $[35]$ and $[45]$ add to
$$\frac{[73][74][26]\langle 6|3+4|5]}{\langle17\rangle\langle27\rangle\langle35\rangle\langle45\rangle\langle12\rangle\langle13\rangle\langle14\rangle\langle26\rangle\langle36\rangle\langle46\rangle\langle56\rangle}$$
But $\langle6|3+4|5] = - \langle6|1+2+7|5]$, by momentum conservation. Making this substitution gives rise to  further terms in $[25], [75]$ and $[15]$. We can first dispose of the $[25]$ term, which is
$$\frac{[73][74][26][25]}{\langle17\rangle\langle27\rangle\langle35\rangle\langle45\rangle\langle12\rangle\langle13\rangle\langle14\rangle\langle36\rangle\langle46\rangle\langle56\rangle}$$
and so antisymmetric in $\{56\}$. The $[75]$ term can be written as
$$\frac{[73][74][75][26]\langle67\rangle\langle34\rangle\langle15\rangle}{\langle12\rangle\langle17\rangle\langle27\rangle\langle35\rangle\langle45\rangle \langle34\rangle \langle13\rangle\langle14\rangle\langle15\rangle \langle26\rangle\langle36\rangle\langle46\rangle\langle56\rangle}$$
where the denominator is antisymmetric in $\{345\}$. Hence symmetrizing over  $\{345\}$ gives zero, again by the three-term Schouten identity.

Hence we are left with terms containing $[15]$, which amount to:
$$\frac{[73][74][15][26]}{\langle12\rangle\langle13\rangle\langle14\rangle\langle27\rangle\langle26\rangle\langle56\rangle\langle15\rangle\langle36\rangle\langle46\rangle\langle34\rangle} \left(\frac{\langle36\rangle \langle14\rangle}{\langle45\rangle\langle73\rangle} - \frac{\langle46\rangle \langle13\rangle}{\langle35\rangle\langle74\rangle} + \frac{\langle15\rangle\langle16\rangle\langle34 \rangle}{\langle17\rangle\langle35\rangle\langle45\rangle}\right)$$
Using $\langle16\rangle\langle34 \rangle = \langle36\rangle \langle14\rangle - \langle46\rangle \langle13\rangle$, this is
\begin{eqnarray}&&\frac{[73][74][15][26]}{\langle12\rangle\langle13\rangle\langle14\rangle\langle27\rangle\langle26\rangle\langle56\rangle\langle15\rangle\langle36\rangle\langle46\rangle\langle34\rangle} \times\nonumber\\&&\left\{\frac{\langle36\rangle \langle14\rangle}{\langle45\rangle}\left(\frac{1}{\langle73\rangle} - \frac{\langle15\rangle}{\langle17\rangle\langle35\rangle}\right) - \frac{\langle46\rangle \langle13\rangle}{\langle35\rangle}\left(\frac{1}{\langle74\rangle} - \frac{\langle15\rangle}{\langle17\rangle\langle45\rangle}\right)\right\}\nonumber \\
&=& \frac{[73][74][15][26]\langle57\rangle\langle67\rangle}{\langle73\rangle\langle74\rangle \langle15\rangle\langle26\rangle\langle 12\rangle\langle27\rangle\langle71\rangle \langle35\rangle\langle45\rangle\langle36\rangle\langle46\rangle\langle56\rangle}\nonumber\\
&=& \frac{\psi^7_3\psi^7_4\psi^1_5\psi^2_6 \langle57\rangle\langle67\rangle}{\langle 12\rangle\langle27\rangle\langle71\rangle\langle35\rangle\langle45\rangle\langle36\rangle\langle46\rangle\langle56\rangle}\end{eqnarray}

On including all the terms from $p=3,4,5,6$ this yields the symmetrized sum which by (\ref{eq:34562}) is equal to
\begin{equation}\frac{\psi^7_{[3}\psi^1_{4}\psi^2_{5}\psi^6_{7]}}{\langle12\rangle \langle26\rangle\langle61\rangle\,\,\langle34\rangle\langle45\rangle\langle53\rangle}\end{equation}
Hence, combining the terms:
\begin{eqnarray}\bar{M}_7(1234567) &=& \frac{\psi^6_{[3}\psi^{1}_{4}\psi^2_{5]}\psi^7_{6}}{\langle12\rangle\langle27\rangle\langle71\rangle \,\,\langle34\rangle \langle45\rangle\langle53\rangle} +  \frac{\psi^3_{[4}\psi^{1}_{5}\psi^2_{6]}\psi^7_{3}}{\langle12\rangle\langle27\rangle\langle71\rangle \,\,\langle64\rangle \langle45\rangle\langle56\rangle}\nonumber\\&+&  \frac{\psi^4_{[3}\psi^{1}_5\psi^2_{6]}\psi^7_{4}}{\langle12\rangle\langle27\rangle\langle71\rangle \,\,\langle36\rangle \langle65\rangle\langle53\rangle}+  \frac{\psi^5_{[3}\psi^{1}_4\psi^2_{6]}\psi^7_{5}}{\langle12\rangle\langle27\rangle\langle71\rangle \,\,\langle34\rangle \langle46\rangle\langle63\rangle}\nonumber\\ &+&  \frac{\psi^7_{[3}\psi^{1}_4\psi^2_{5}\psi^6_{7]}}{\langle12\rangle\langle26\rangle\langle61\rangle \,\,\langle34\rangle \langle45\rangle\langle53\rangle}\label{eq:7graviton}\end{eqnarray}

But $\bar{M}_7$ is symmetric in its arguments. Take the symmetric part in $\{712\}$ and obtain:
\begin{eqnarray}\bar{M}_7(1234567) &=& \frac{1}{3}\{R(712345;6) + R(456712;3)+ R(567123;4) + R(671234;5)\nonumber\\&+& R(123456;7) + R(234567;1)+R(345671;2)\}\end{eqnarray}
as required. 

The formula (\ref{eq:7graviton}) gives a shorter expression (in 42 terms) for $\bar{M}_7$, but not showing its symmetry so clearly. In contrast, expansion of the BGK formula will give rise to a sum of 96 simple products of spinors. It is reasonable to suppose that this economy can be extended to higher $n$. Our
 results also indicate that there is rich algebraic structure in these expressions, and suggest that a new normal form for the $n$-graviton MHV amplitude may be constructed out of the anti-symmetrized phase factors.
\addtocounter{equation}{+1}
\newpage
\section{NMHV amplitudes}

In this section we will evaluate the NMHV amplitude $M(1^+2^+3^+4^-5^-6^-)$ on the assumption that the N=7 BCFW recursion is valid.\footnote{Section 13, added in version 2, shows why this assumption is indeed justified.} We take $\{3^+4^-\}$ as the pivots. The terms arising from BCFW are as follows:

First, one term from $M(3^+5^-6^-k)\circ M(4^-1^+2^+k)$, with diagram
 \begin{figure}[h] 
   \centering
   \includegraphics[width=163px]{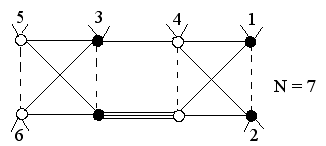} 
  \end{figure}
\vspace{-25mm}
\begin{equation} \end{equation}
\vspace{5mm}

We will refer to this as $M_{356}$, characterising it by its factor of $s_{356}$.

Next, the term $M(3^+1^+5^-k)\circ M(4^-6^-2^+k),$ with diagram
 \begin{figure}[h] 
   \centering
   \includegraphics[width=161px]{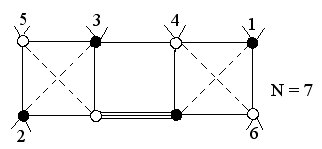} 
  \end{figure}
\vspace{-25mm}
\begin{equation} \end{equation}
\vspace{5mm}

and three similar terms obtained by $1\leftrightarrow2, 5\leftrightarrow6.$
We can refer to these as $M_{315}, M_{325}, M_{316}, M_{326}$.

Then there is the term $M(3^+1^+5^-6^-k)\circ M(4^-2^+k), $ with a diagram which has to be written as a sum:
 \begin{figure}[h] 
   \centering
   \includegraphics[width=244px]{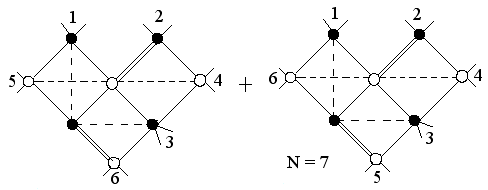} 
  \end{figure}
  \vspace{-25mm}
\begin{equation} \end{equation}
\vspace{10mm}

There is a second term obtained by  $1 \leftrightarrow2,$ and the `flipped' versions of these, i.e.\ those obtained by the mapping $(123456)\leftrightarrow(654321), \langle\leftrightarrow[$.
We will refer to these as $M_{342}, M_{341}, M_{345}, M_{346}$. 

The amplitude is then the sum of these nine terms.
 (There are ten ways of dividing the six momenta into two groups of three, but the tenth corresponds to $M_{123}=0$.) Only three types of term need to be considered.

Explicitly, evaluating the diagrams, we have
\begin{eqnarray}M_{356} &=& \frac{[12]^7\langle 56\rangle^7\langle12\rangle [56]}{s_{356} [41][42]\langle35\rangle\langle 36 \rangle  [1|2+4|3\rangle [2|1+4|3\rangle \langle5|1+2|4]\langle 6|1+2|4]}\\
M_{315}&=& \frac{[2|4+6|5\rangle^7 \langle26\rangle [15]}{s_{315}[24][26][46]\langle13\rangle \langle15\rangle \langle35\rangle \langle1|3+5|4] \langle 3|1+5|6] \langle 3|1+5|4]}\end{eqnarray}

and $M_{316}, M_{325}, M_{326}$ are similar, by $1\leftrightarrow 2$ and/or $5\leftrightarrow 6.$ Lastly,
\begin{eqnarray}M_{342}&= &\frac{\langle 4|2+3|1]^7 [24]}{s_{342}[56]\langle24\rangle\langle34\rangle\langle23\rangle[1|2+4|3\rangle [5|2+4|3\rangle [6|2+4|3\rangle}\nonumber \\ &\times &\left\{\frac{\langle15\rangle\langle36\rangle}{[15]\langle2|3+4|6]} - \frac{\langle16\rangle\langle35\rangle}{[16]\langle2|3+4|5]} \right\}\end{eqnarray}
and then $M_{341}$ is found by $1 \leftrightarrow 2$; $M_{345}, M_{346}$ by $(123456)\leftrightarrow (654321), \langle \leftrightarrow [$.

The spurious pole structure is as follows. $M_{412}$ has four spurious poles. $M_{426}, M_{416},$ $ M_{425}, M_{415}$ have only three, and $M_{341}, M_{342}, M_{345}, M_{346}$ have five. There are 18 different spurious poles, and each of them occurs in just two terms. For instance, $\langle2|3+4|5]$ occurs in the $M_{342}$ and $M_{345}$ terms, and demands that the combination
\begin{equation}\frac{[24]\langle35\rangle}{\langle2|3+4|5]}\left\{\begin{array}{r}\displaystyle{\frac{\langle 4|2+3|1]^7\langle16\rangle}{s_{342}[16][56]\langle24\rangle\langle34\rangle\langle23\rangle[1|2+4|3\rangle [5|2+4|3\rangle [6|2+4|3\rangle}}\\\vspace{0mm}\\ \displaystyle{-\frac{\langle 6|4+5|3]^7[16]}{s_{345}\langle16\rangle\langle12\rangle[35][34][45][4|3+5|6\rangle [4|3+5|2\rangle [4|3+5|1\rangle}}\end{array}\right\}\end{equation}
is non-singular when $\langle2|3+4|5]=0$. The author is grateful to Jacob Bourjaily for checking numerically with his powerful {\sc Mathematica} application (Bourjaily 2010) that this is indeed the case.

We may also check the asymptotic behaviour under momentum shifting. Analogously with the discussion of the MHV amplitudes, the shift we use is the shift used for $\{3^+4^-\}$ without super-symmetry, namely
\begin{equation}4] \rightarrow 4] + z\, 3]; \quad 3\rangle \rightarrow 3\rangle - z\,4\rangle\, .\end{equation}
In the $M_{356}$ term the numerator is actually invariant under the shift, whilst each factor in the denominator is linear in $z$,  so this term is of behaviour $z^{-9}$.

In the terms like $M_{342}$, the numerator factor $\langle4|2+3|1] = \langle42\rangle[21] + \langle43\rangle[31]$ is again invariant under the shift. The numerator $[24]$ contributes linearly in $z$, and so do $\langle36\rangle$ and $\langle35\rangle$. In the denominator there are six factors linear in $z$, so the overall behaviour is like $z^{-4}$.

So if the nine terms add up to $M(1^+2^+3^+4^-5^-6^-)$, it must be that $z^{-2}$ behaviour is found in the $M_{315}$ terms. Here the numerator factor $[2|4+6|5\rangle^7$ contributes a $z^7$ power. But the denominator has $z^9$ behaviour, when proper account is taken of the fact that $\langle3|1+5|4]$ becomes {\em quadratic} in $z$. Hence the overall behaviour of these four terms is indeed  like $z^7/z^9 = z^{-2}$. Again, therefore, we find that each one of the nine terms conforms to the $z^{-2}$ behaviour of the overall amplitude.

The nine-term expression is simpler, though not hugely simpler, than that given by Cachazo and  Svr\u{c}ek (2005). It has fewer terms, no double poles, and fewer spurious poles;  it follows directly from the recursion rules without needing intermediate simplification of the sub-amplitude formulas. As a nine-term expression it is not excessively long, considering that $60\times 3$ terms are needed for the complete gauge-theoretic NMHV amplitude, with all colour orders. But of course we are very far from seeing the $S_6$ symmetry of the full super-amplitude, and we have no real insight into the spurious pole structure. In the gauge-theoretic analogue, representation by {\em momentum twistors}, as introduced in (Hodges 2009), has cast new light on the spurious poles. But in the gravitational case it is not clear how we might use such a representation effectively. In any case, we have no hope of understanding the geometry of the NMHV amplitudes unless the MHV amplitudes are first made clear.

One feature of the N=7 calculus is still notably simple: all these diagrams involve simple {\em numerators} integrated over super-twistor regions. This simplicity is completely concealed when the amplitudes are written out in momentum-space.\footnote{{\em Note added for Version 2:} iIn the light of the general justification of the N=7 recursion, added in section 13,  we can enjoy greater confidence that the advantageous features of this representation of the six-field NMHV amplitude will extend to all tree amplitudes.} 

 The foregoing remarks serve to motivate the last development, in which we write down the amplitudes in momentum-twistor space and study the emergence of numerator factors.

\newpage

\section{The momentum-twistor numerator}

In this final section we will turn to a further development, in which we investigate the representation in momentum-twistor space of the gravitational MHV amplitudes. This development is logically independent of the preceding material, since we are not concerned now with how the amplitude is computed. However, it is partly motivated by the structure elucidated by the N=7 diagrams, in which an integrand numerator  appears to express the conformal-symmetry-breaking feature of gravity. Can momentum-twistor space provide a formulation in which a simple numerator appears in the statement of the amplitude?

The essence of the gravitational setting is that we do not have any ring-ordering of the external fields, whilst the definition of momentum-twistors requires an ordering. We will adopt the point of view that the resulting structure should be {\em order-independent,} and that this independence will encode the $S_n$ symmetry.

As a precursor, we consider the  gravitational soft factor, as  discussed by Nguyen, Spradlin, Volovich and Wen (2009). This can be given by the formula:
\begin{equation}\sum_{j=1}^{n-1} \frac{[jn]\langle j \alpha\rangle\langle j\beta\rangle}{\langle jn\rangle\langle n\alpha\rangle\langle n\beta\rangle}= \sum_{j=1}^{n-1} \frac{ \langle j \alpha\rangle\langle j\beta\rangle}{\langle n\alpha\rangle\langle n\beta\rangle}\psi^j_n \label{eq:soft} \end{equation}

Here momentum conservation implies that the sum is independent of the choice of $\alpha$ and $\beta$.
In particular we may choose $\alpha=n-1, \beta =1$.

As the invariance of this expression depends upon momentum conservation, this suggests that momentum-twistor variables, in which this conservation is automatically encoded, may give it a simple form. In fact, if we take momentum twistors with respect to the order $\{1234\ldots n-1, n\}$, we find that 
for $n\ge 4$, the soft factor is economically expressed by:
\begin{equation}\sum_{j=2}^{n-2} \frac{\langle n-1,n1j\rangle\langle j-1,j+1\rangle}{\langle1n\rangle\langle n-1,n\rangle\langle j-1,n\rangle\langle jn\rangle\langle j+1,n\rangle}\, .\end{equation}

This is encouraging, and leads us to investigate the form taken by the full gravitational MHV amplitude in such co-ordinates.

As usual we begin with the $n=3$ amplitude, and note that instead of thinking of it as the {\em square} of the gauge-field amplitude, we may write it as:
$$ \frac{1}{\prod_i \langle i, i+1\rangle \,\, \prod_{i < j} \langle ij\rangle}$$
Considering next the $n=4$ amplitude, we may take momentum-twistors defined relative to the order (1234), and find that it also fits this pattern:
\begin{equation}\tilde{M}_4(1234) =  \frac{\langle 1234\rangle}{\prod_i \langle i, i+1\rangle \,\, \prod_{i < j} \langle ij\rangle}\end{equation}
The notation $\tilde{M}$ has been used because this is not the same as the reduced amplitude $\bar{M}$; the arguments are momentum twistors rather than pairs of spinors, and it is no longer the co-efficient of the delta-function in momenta.

The denominator factor $\prod_i \langle i, i+1\rangle$ is essentially  a Jacobian, giving the effect of translating into momentum-twistor space. The other denominator simply gives the product of all the possible physical poles.  The numerator $\langle1234\rangle$  therefore captures the essential content of the 4-graviton amplitude.

It is not obvious that the five-field amplitude will take any analogous form. We have the expression (\ref{eq:mbar5})
in which the denominator is already given as $\prod_{i < j} \langle ij\rangle$.
Encoding the numerator with momentum-twistors defined relative to (12345) we find, after a little algebra,
\begin{equation}\tilde{M}_5(12345) = \frac{\langle1234\rangle\langle2345\rangle\langle51\rangle - \langle1234\rangle\langle3451\rangle\langle25\rangle - \langle5123\rangle\langle2345\rangle\langle14\rangle\,}{\prod_i \langle i, i+1\rangle \,\, \prod_{i < j} \langle ij\rangle} \, .\end{equation}
The remarkable thing about this expression is that it is indeed a pure numerator, i.e.\ a polynomial $N_5(12345)$ of degree 2 in each of  the five momentum twistors. The two terms in (\ref{eq:mbar5}) do not individually have this property. 

It is even less obvious that the six-field amplitude will have a similar form. Insertion of the BGK formula  leads to some very unpromising expressions. However, we may craftily choose the particular representation
\begin{equation}\bar{M}_6(123456) =  \frac{\psi^1_{[4}\psi^3_6\psi^5_{2]}}{\langle13\rangle \langle35\rangle\langle51\rangle\,\,\langle46\rangle\langle62\rangle\langle26\rangle}\, ,\end{equation}
which translates immediately into:
\begin{equation}\tilde{M}_6(123456) = \frac{N_6(123456)}{\prod_i \langle i, i+1\rangle \,\, \prod_{i < j} \langle ij\rangle}\end{equation}
where 
\begin{eqnarray}N_6(123456)&= &\langle 123I456\rangle \langle 234I561 \rangle \langle 345I612\rangle\nonumber\\
&+& \langle  123I456 \rangle \langle5612\rangle\langle 2345\rangle\langle14\rangle\langle36\rangle \nonumber\\
&+& \langle  234I561 \rangle \langle6123\rangle\langle 3456\rangle\langle25\rangle\langle41\rangle\nonumber\\
&+& \langle  345I612 \rangle \langle1234\rangle\langle 4561\rangle\langle36\rangle\langle52\rangle \nonumber\\
&+& \langle 1234\rangle \langle3456\rangle\langle5612\rangle \langle 14\rangle\langle25\rangle\langle36\rangle\nonumber\\ 
&+& \langle 2345\rangle \langle4561\rangle\langle6123\rangle \langle 25\rangle\langle36\rangle\langle41\rangle \, . 
\end{eqnarray}
Here  $\langle123I456\rangle$ represents $\epsilon_{\alpha\beta\gamma\kappa}Z_1^{\alpha}Z_2^{\beta}Z_3^{\gamma}I^{\kappa\lambda}Z_4^{\mu}Z_5^{\nu}Z_6^{\pi}\epsilon_{\lambda\mu\nu\pi}$.

The existence of this polynomial numerator thus seems to be associated with the new representations of the MHV amplitudes found in section 10. So the conjecture made in that section about the extension to general $n$ of the new representation of amplitudes, suggests a further conjecture put in terms of momentum twistor space.

First we define what is meant by order-independence, in a way that does not depend on translating back into the original spinors but is intrinsic to momentum-twistor space.

Let $\sigma$ be  a permutation on $(123\ldots n)$, and let $(1'2'3'\ldots n')$ denote the momentum twistors as defined relative to the ordering $\sigma(1), \sigma(2), \sigma(3), \ldots \sigma(n)$. Then $F_n(123\ldots n)$ will be said to be order-independent if for all $\sigma$.
\begin{equation}\frac{F_n(1'2'3'\ldots n')}{\prod_i \langle\sigma(i)\sigma(i+1)\rangle} = \frac{F_n(123\ldots n)}{\prod_i \langle i, i+1\rangle}\, .\end{equation}
Of course it not necessary to check all $\sigma$. It is necessary and sufficient (i) that  $F$ is {\em cyclic}, and (ii) that 
$F$ transforms correctly under a transposition of just one pair of neighbouring elements in the chosen ordering. Under such a transposition, all but two of the momentum twistors may remain unchanged, and there is a simple formula for the other two. Explicitly, suppose we compare the momentum twistors $Z_i^{\alpha}$ as defined by the order $(1234\ldots n)$ with the $Z'^{\alpha}_i$ as defined by $(1324\ldots n)$. The $Z'^{\alpha}_i$ can be taken to be the same as the $Z^{\alpha}_i$ except for:
\begin{equation} Z'^{\alpha}_2 = (Z_3^{\alpha}\langle42\rangle - Z_4^{\alpha}\langle32\rangle)\langle34\rangle^{-1}, \,\, Z'^{\alpha}_3 = (Z_2^{\alpha}\langle31\rangle - Z_1^{\alpha}\langle32\rangle)\langle12\rangle^{-1}\, \end{equation} Thus it is sufficient if
$\langle12\rangle\langle34\rangle F_n(12'3'\ldots n) = - \langle13\rangle\langle24\rangle F_n(123\dots n)$.

It is easy to see that $\langle1234\rangle$ satisfies this criterion.

{\bf Conjecture}: 
\begin{equation}\tilde{M}_n(123\ldots n) = \frac{N_n(123\ldots n)}{\prod_i \langle i, i+1\rangle \,\, \prod_{i < j} \langle ij\rangle}\end{equation}
where $N_n$ is a polynomial of degree $(n-3)$ in each of the $n$ momentum twistors, and of degree $(n-3)(n-4)/2$ in $I$, order-independent in the sense defined above.

If this conjecture is correct, it implies some non-obvious property satisfied by the MHV amplitudes. We would expect this to be associated with some new geometric structure. In the case of the five-graviton numerator  $N_5$, there is a structure which manifests its cyclicity. This is because for $i=1,2,3,4, 5,$ we may define 
$z_i^{A'} = \langle i+1, i+2, i+3, i+4\rangle\pi_i^{A'}$, where $Z_i^{\alpha} = (\omega_i^{A'}, \pi_{iA'})$. Then by the 5-term Schouten identity on twistors, $z_1^{A'} +  z_2^{A'}+ z_3^{A'}+ z_4^{A'}+ z_5^{A'}=0$ in ${\mathbb C}^2$. 
These five vectors therefore form a closed path on ${\mathbb C}^2$, which we may draw thus:
\newpage
 \begin{figure}[h] 
   \centering
   \includegraphics[width=202px]{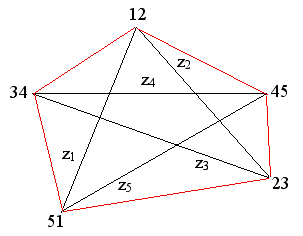} 
  \end{figure}
    \vspace{-25mm}
\begin{equation} \end{equation}
\vspace{5mm}

The quantity $N_5 = z_1.z_5 + z_2.z_5 + z_1.z_4$ then has a geometrical meaning as the (twice) the (complexified) area of a pentagon, {\em not} the pentagon defined by the given five edges, but the (red) pentagon with vertices (12), (34), (51), (23), (45) in order. It is thus manifestly cyclic. 

This pentagonal area is reminiscent of similar expressions in the  recent geometric description of amplitudes (Arkani-Hamed, Bourjaily, Cachazo, Hodges and Trnka 2010). It remains to be seen whether such geometrical pictures can be extended to show the full symmetry of the $n$-graviton MHV amplitudes.

\section{Additional material for Version 2}
\subsection{Justification of N=7 recursion in general}

The behaviour shown by the 4-field amplitudes compared in section 7 is in fact generic, and this fact is sufficient to justify the N=7 recursion for all tree-level amplitudes. The author is indebted for this point to Henriette Elvang, who has referred to (Elvang, Huang and Peng, 2011) for the demonstration.  Section 4 of that paper establishes the  simple and powerful statement that a shifted gauge-theoretic amplitude, using {\em any} shift defined by  adjacent pivots, will exhibit $z^{3-N}$ behaviour. Hence N=3 super-symmetry will fail to give rise to a valid BCFW recursion. But Henriette Elvang points out that the argument used in that paper applies equally to super-gravity and shows a $z^{6-N}$ behaviour for any shifted amplitude. Hence any N=7 super-gravity amplitude will necessarily have $z^{-1}$ behaviour, sufficient to justify the BCFW decomposition.  

The shifted super-momenta, as employed in these arguments, give rise to BCFW expressions which correspond directly to  the super-twistor diagrams of this paper. This connection is explained in (Arkani-Hamed, Cachazo,  Cheung and  Kaplan, 2009a). It has a simple origin: the original non-super-symmetric diagram theory (Hodges 2005a) showed how the twistor diagrams and their composition rule were equivalent to the non-super-symmetric BCFW terms and their composition rule. This connection is preserved when super-symmetric extension is applied to both formalisms. Thus the results of (Elvang, Huang and Peng, 2011) apply directly to the N=7 diagrams used in this paper.  A simple induction on $n$, starting with $n=3$, then shows that the  BCFW decomposition rule using N=7 diagrams is valid for all $n$-graviton tree amplitudes of all helicities.

Of course, this argument only shows that the N=7 BCFW recursion is valid, and does not show how or why it has the special features of simplicity that have been explored above. 

\subsection{Further addenda and corrigenda}

This Version 2 omits the equation (79) stated in version 1, which was erroneous. The author is indebted to Yu-Xiang Gu (Perimeter Institute) for checking the claim and finding the error. Fortunately, no other statements in the paper rested on this assertion.

The author is also grateful to Howard Schnitzer for pointing out the earlier interesting and profitable use made of momentum-twistors in a super-gravity context  in (Naculich, Nastase and Schnitzer 2011).

\section{Acknowledgements}

The  N=7 recursion was first formulated during my May-June 2010 visit to the Institute for Advanced Study, thanks to an invitation from Nima Arkani-Hamed, and I am most grateful  for generous support from that institution. A second visit in May  2011 allowed  further highly stimulating discussions of this material and specific suggestions from Nima Arkani-Hamed. The interest   and encouragement of Marcus Spradlin and Anastasia Volovich has also been notable, and I am grateful for the invitation to join them in April 2011 at the  Kavli Institute for Theoretical Physics, UC Santa Barbara, for the Harmony of Scattering Amplitudes workshop. At Oxford, I am particularly grateful to  Lionel Mason and Tsou Sheung Tsun for organizing  the special conference in July 2011 marking the eightieth birthday of Sir Roger Penrose, FRS, OM. The  influence of Roger Penrose's ideas, over 40 years, remains paramount.
\newpage

\section{References}

N. Arkani-Hamed, F. Cachazo, C. Cheung and J. Kaplan, The S-matrix in twistor space, arXiv:0903.2110 (2009a)

N. Arkani-Hamed, F. Cachazo, C. Cheung, and J. Kaplan, A duality for the S-matrix, arXiv:0907.5418 (2009b)

N. Arkani-Hamed, J. Bourjaily, F. Cachazo, A. Hodges and J. Trnka,   A note on polytopes for scattering amplitudes, arXiv:1012.6030 (2010)

N. Arkani-Hamed, personal communication and  presentation at the \newline RP80 Oxford conference on {\em Twistors, Geometry and Physics,} \newline http://people.maths.ox.ac.uk/lmason/rp80.html (2011)

J. Bedford, A. Brandhuber, W. Spence and G. Travaglini, A recursion relation for gravity amplitudes, arXiv:hep-th/0502146v2 (2005)

F. A. Berends, W. T. Giele and H. Kuijf, Phys.\ Lett.\ B  {\bf211}, 91 (1988).

J. L. Bourjaily, Efficient tree-amplitudes in N=4: automatic BCFW recursion in {\sc Mathematica}, arXiv:1011.2447 (2010)

R. Britto, F.  Cachazo, B. Feng and E. Witten, Phys.\ Rev.\ Lett.\ {\bf 94}, 181602 
[arXiv:hep-th/0501052] (2005)

J. J. Carrasco and H. Johansson, Five-point amplitudes in N=4 super-Yang-Mills theory and N=8 supergravity, arXiv:1106.4711 (2011)

F. Cachazo and P. Svr\u{c}ek, Tree level recursion relations in general relativityC, arXiv:hept-th/0502160v3  (2005)

B. S. DeWitt, Quantum theory of gravity. III. Applications of the covariant theory, Phys.\ Rev.\ {\bf 162,} 1239 (1967)

J. M. Drummond, M. Spradlin, A. Volovich and C. Wen, Tree-level amplitudes in N=8 supergravity, arXiv:0901.2363v2 (2009)

H. Elvang, Y.\ t.\ Huang and C. Peng, On-shell superamplitudes in N $<$ 4 SYM, arXiv:1102.4843v1 (2011)

A. Hodges, Twistor diagram recursion for all gauge-theoretic tree amplitudes,  \newline  arXiv:hep-th/0503060  (2005a)

A. Hodges, Twistor diagrams for all tree amplitudes in gauge theory: a helicity-independent formalism, arXiv:hep-th/0512336 (2005b)

A. Hodges, Scattering amplitudes for eight gauge fields, arXiv:hep-th/0603101 (2006)

A. Hodges, Eliminating spurious poles from gauge-theoretic amplitudes, \newline  arXiv:0905.1473 (2009)

H. Kawai, D. C. Lewellen and S. H. H. Tye, Nucl.\ Phys.\ B {\bf 269,} {1} (1986)

L. J. Mason and D.  Skinner, Gravity, twistors and the MHV formalism, \newline arXiv:0808.3907v2 (2009)

S. G. Naculich, H. Nastase and H. J. Schnitzer, Applications of subleading color amplitudes in N = 4 SYM theory, arXiv:1105.3718

D. Nguyen, M. Spradlin, A. Volovich, and C. Wen, The tree formula for MHV graviton amplitudes, arxiv:0907.2276v2 (2009)

R. Penrose and M. A. H. MacCallum, Twistor theory: an approach to the quantisation of fields and space-time, Physics Reports {\bf 4,} 241 (1972) 

R. Penrose, Twistor theory, its aims and achievements, in {\em Quantum Gravity,} eds. C. J. Isham, R. Penrose and D. W. Sciama, Oxford University Press (1975)

 R. Penrose, Non-linear gravitons and curved twistor theory, Gen.\ Rel.\ Grav. {\bf 7},  31 (1976)

\end{document}